\documentclass[a4paper,twocolumn,11pt,accepted=2024-01-03]{quantumarticle}
\pdfoutput=1
\usepackage[utf8]{inputenc}
\usepackage[english]{babel}
\usepackage[T1]{fontenc}
\usepackage{amsmath}
\usepackage{hyperref}

\usepackage{tikz}
\usepackage{lipsum}

\usepackage{siunitx}
\usepackage{graphicx}
\usepackage{subfig} 
\graphicspath{ {images/} }
\usepackage{subfig} 
\usepackage{amsthm}
\usepackage{cite}

\usepackage{appendix}
\usepackage{amssymb}
\usepackage{secdot}
\usepackage[T1]{fontenc}
\usepackage{indentfirst}
\usepackage{color}
\usepackage{braket}
\usepackage{bbold}
\usepackage{dsfont}
\usepackage{multirow}
\usepackage{rotating}
\usepackage[newcommands]{ragged2e}
\usepackage{graphicx,caption}
\usepackage{listings}
\DeclareMathOperator{\diag}{diag}

\usepackage{xcolor,colortbl}

\definecolor{Gray}{gray}{0.85}
\definecolor{LightCyan}{rgb}{0.88,1,1}
\definecolor{applegreen}{rgb}{0.55, 0.71, 0.0}
\definecolor{carnationpink}{rgb}{1.0, 0.65, 0.79}

\theoremstyle{definition}

\newtheorem{observation}{Observation}
\newtheorem{proposition}{Proposition}
\newtheorem{theorem}{Theorem}

\newtheorem{conjecture}{Conjecture}

\usepackage{cleveref}
\crefname{proposition}{Proposition}{Propositions}
\crefname{definition}{Definition}{Definitions}
\crefname{lemma}{Lemma}{Lemmas}
\crefname{figure}{Fig.}{Fig.}
\crefname{corollary}{Corollary}{Corollary}
\crefname{conjecture}{Conjecture}{Conjectures}
\crefname{section}{Section}{Sections}
\crefname{appendix}{Appendix}{Appendixes}
\crefname{observation}{Observation}{Observation}
\crefname{remark}{Remark}{Remark}
\crefname{example}{Example}{Examples}
\crefname{equation}{}{}
\crefname{table}{Table}{Tables}
\crefname{section}{Section}{Sections}

\begin{document}

\title{Orthonormal bases of extreme quantumness}

\author{Marcin Rudzi{\'n}ski}
\email{marcinwojciechrudzinski@gmail.com}
\orcid{0000-0002-6638-3978}
\affiliation{
Faculty of Physics, Astronomy and Applied Computer Science, Jagiellonian University, ul. \L{}ojasiewicza 11, 30-348 Krak{\'o}w, Poland}
\affiliation{Doctoral School of Exact and Natural Sciences, Jagiellonian University, ul. \L{}ojasiewicza 11, 30-348 Krak{\'o}w, Poland}

\author{Adam Burchardt}
\email{adam.burchardt.uam@gmail.com}
\orcid{0000-0003-0418-257X}
\affiliation{QuSoft, CWI and University of Amsterdam, Science Park 123, 1098 XG
Amsterdam, the Netherlands}

\author{Karol {\.Z}yczkowski}
\email{karol.zyczkowski@uj.edu.pl}
\orcid{0000-0002-0653-3639}
\affiliation{
Faculty of Physics, Astronomy and Applied Computer Science, Jagiellonian University, ul. \L{}ojasiewicza 11, 30-348 Krak{\'o}w, Poland}
\affiliation{Center for Theoretical Physics, Polish Academy of Sciences,  Al. Lotnik{\'o}w 32/46, 02-668 Warszawa, Poland}

\maketitle

\begin{abstract}
  Spin anticoherent states acquired recently a lot of attention as the most "quantum" states. Some coherent and anticoherent spin states are known as optimal quantum rotosensors. In this work, we introduce a measure of quantumness for orthonormal bases of spin states, determined by the average anticoherence of individual vectors and the Wehrl entropy. In this way, we identify the most coherent and most quantum states, which lead to orthogonal measurements of extreme quantumness. Their symmetries can be revealed using the Majorana stellar representation, which provides an intuitive geometrical representation of a pure state by points on a sphere. Results obtained lead to maximally (minimally) entangled bases in the $2j+1$ dimensional symmetric subspace of the $2^{2j}$ dimensional space of states of multipartite systems composed of $2j$ qubits.  Some bases found are iso-coherent as they consist of all states of the same degree of spin-coherence.
\end{abstract}

\section{Introduction}

Geometric methods play an essential role while studying physical systems in classical mechanics \cite{Frankel,03}, relativity \cite{06}, quantum mechanics \cite{Zyczkowski,05}, and quantum field theory \cite{04}. The stellar representation, also called \textit{the Majorana representation}, is one of the important geometrical representations in quantum mechanics \cite{Majorana}. The stellar representation presents a spin-$j$ pure state in $N=2j+1$ dimensional Hilbert space as a collection of $2j$ points on a sphere. The same constellation represents a symmetric state of a system consisting of $2j$ qubits. In the case of a spin-$\frac{1}{2}$ particle, it reduces to the celebrated Bloch representation of a two-level quantum system (qubit).  
This representation is used in various contexts such as spinor Bose gases \cite{Barnett1,Barnett2,Makela,Ensastiga}, entanglement classification in multiqubit systems \cite{Mathonet, MartinMultiqubit,Aulbach,Markham,Ribeiro,Aulbach2,Ganczarek,Mandilara,Hyllus}, the Berry phase associated with the cyclic evolution of the state \cite{Hannay,Bruno,Liu}, investigating Lipkin-Meshkov-Glick model \cite{Ribeiro1,Ribeiro2} and studying symmetries and properties of spin states \cite{Baguette-multiqubit,Baguette-point-group,Zimba,GiraudTensor,CohMAjo,Baguette-measure,Kolenderski,Chryssomalokos,GoldbergOptimal,MartinOptimal,Crann,Bannai,Wang,Grassl, Goldberg,GiraudQueens}.

The Majorana representation appears naturally in the context of $SU(2)$ coherent and anticoherent states of a given spin $j$ \cite{Baguette-multiqubit,Baguette-measure,GiraudTensor,Baguette-point-group,Zimba,CohMAjo}. Note that the spin coherence is not basis dependent. Properties of a coherent state $\ket{\mathbf{n}}$ of size $2j+1$ closely resemble the classical state of spin $j$ pointing in the direction given by the vector $\mathbf{n}$. It has minimal uncertainty of spin operator $\mathbf{S}$ \cite{Delbourgo} and its Wehrl entropy \cite{Wehrl1979,Lieb_conjecture} attains the minimal value \cite{Lee1988,Lieb_proof}.  In the Majorana representation the most coherent state is represented by one $2j$-degenerated point on a sphere. On the other hand, the most quantum, or the most anticoherent state $\ket{\psi}$ should "point nowhere" and be represented as $2j$ points equally distributed on a sphere \cite{Grassl, Goldberg,GiraudQueens}. However, even if the polarization (coherence) disappears, the higher moments of coherence might not vanish. Hence,  anticoherence can be defined up to a given order \cite{Zimba}. Highly anticoherent states turned out to be applicable as optimal quantum rotosensors \cite{Kolenderski,Chryssomalokos,GoldbergOptimal,MartinOptimal}. They also coincide with spherical $t$-designs in several dimensions \cite{Crann,Wang}, however, a general conjecture concerning their relation was disproved \cite{Bannai}. Experimental realization of some of those states was discussed in \cite{Bouchard}. The Majorana representation has proven to be a suitable tool to study properties of anticoherent states. 

As much as the construction of coherent states is straightforward for $N>2$, it is not possible to construct an orthonormal basis composed only of coherent states. Furthermore, much effort has been made to determine quantum states with the highest possible quantumness or anticoherence, while the concept of the most anticoherent, orthonormal basis of states is largely unexplored. States of extreme quantumness are often optimal in several measurement scenarios. Such bases can thus provide an effective tool to extend the range of the measured parameter or for multiparameter estimations, for example, using the Bayesian analysis.

The main goal of this paper is to address the problem of finding orthogonal bases of extremal properties in an $N$-dimensional Hilbert space.
On one hand, we look for
the most spin-coherent bases, which provide "the most classical" quantum orthogonal measurement.
On the other hand, we aim to identify the quantum measurement  "as quantum as possible", composed of orthogonal vectors which are the least spin-coherent and maximize the average measure of anticoherence.
Several quantities can be used for this purpose, including the Wehrl entropy, which characterizes localization of the state in the phase space \cite{We78, We91, Gz01, Kz01, Wehrl1979, Lieb_conjecture, Lieb_proof, Lee1988}, measures related to the distribution of Majorana stars representing a state \cite{Ganczarek} or cumulative distribution based on multipole expansion of density matrix \cite{Grassl,LLSS}.
In this work we rely on the quantity introduced by Baguette and Martin  \cite{Baguette-measure}, under the name of measure of anticoherence.

To highlight the symmetries that arise in the studied quantum structures, we use the Majorana stellar representation. The constellations representing the bases and vectors obtained in this study exhibit classical symmetries, such as a Platonic solid, its compound or an Archimedean solid. Geometric configurations of Platonic solids appear in many areas of quantum information theory. In particular, such structures were recently used to construct particular classes of quantum measurements \cite{Tavakoli,Nguyen}, absolutely maximally entangled (AME) states \cite{Latorre} and Bell inequalities \cite{Bolonek,Pal}.

This work is organized as follows. \cref{Stellar_representation} recalls the Majorana representation for a spin-$j$ state $\ket{\psi}\in\mathcal{H}_{2j+1}$ or for a symmetric state of $2j$ qubits. \cref{Anticoherence_measures} presents a measure of quantumness of orthonormal bases, defined using an anticoherence measure initially introduced for quantum states \cite{Baguette-measure}. \cref{Main_results} describes our methods of searching for the most "classical" and the most "quantum" orthonormal bases and presents the results obtained. In \cref{Wehrl} we confront our results with a measure of quantumness given by the mean Wehrl entropy of vectors forming the basis and the maximum of the Husimi function. Bases of extreme entanglement in the symmetric subspace of $2j$-qubit systems are discussed in \cref{Permutation-sym-states}.

\section{The stellar representation}
\label{Stellar_representation}

The Bloch sphere is a geometrical representation of pure states of a two-level quantum system, often called a qubit. In particular, a state of spin-$\frac{1}{2}$ quantum system, can be naturally represented as a point on the sphere. Indeed, a pure quantum state 
\begin{equation}
\label{eq1}
Z_0 \ket{\tfrac{1}{2},\tfrac{1}{2}} + 
Z_1 \ket{\tfrac{1}{2},-\tfrac{1}{2}},
\quad\quad
|Z_0|^2 +|Z_1|^2 =1
\end{equation} 
uniquely determines a complex number $z=Z_1/Z_0$ (possibly $z=\infty$), which can be projected onto the surface of a 2-dimensional sphere by the stereographic projection:
\begin{equation}
\label{stereographicProj}
z\mapsto 
\big(\theta, \phi \big):= \big(2 \text{ arctan} |z| \, , \, \text{arg} (z) \big),
\end{equation}
where $\theta \in [0,\pi]$ and $\phi \in [0,2\pi ]$ are usual spherical coordinates. Note that the stereographic projection provides a bijection between extended complex plane $\mathbb{C}\cup\{\infty\}$ and the surface of the sphere, the reverse transformation is given by $z=e^{i\phi}\tan(\theta/2)$. Furthermore, a complex number $z$ uniquely defines the quantum state \cref{eq1} satisfying the property $z=Z_1/Z_0$. 

In 1932 Ettore Majorana generalized the celebrated Bloch representation for arbitrary spin-$j$ states \cite{Majorana}. The \textit{stellar representation} maps a spin-$j$ state $\ket{\psi}$ from $(2j+1)$-dimensional Hilbert space $\mathcal{H}_{2j+1}$ into a constellation of $2j$ points on a sphere. More precisely, any spin-$j$ state might be expressed in the basis of the angular momentum operator $J_z$ as
\begin{equation}
    \ket{\psi}=\sum_{m=-j}^{j}Z_{j-m}\ket{j,m} \;\in\; \mathcal{H}_{2j+1},
    \label{state-ang-mom-bas}
\end{equation}
where $\sum_{m=-j}^j|Z_{j-m}|^2=1$. The state coefficients are further used to construct \textit{the Majorana polynomial} of degree $2j$ in complex variable $z$,
\begin{equation}
    w(z)=\sum_{k=0}^{2j}(-1)^k Z_k\sqrt{\binom{2j}{k}}z^{2j-k}.
    \label{polynomial}
\end{equation}
Such a polynomial uniquely determines $2j$ possibly degenerated roots $z_i$, $i=1,\ldots, 2j$, which can be mapped on a sphere by the stereographic projection \cref{stereographicProj}. In this way, a state \cref{state-ang-mom-bas} is represented as $2j$ points on a  sphere, which are called \textit{stars}. We refer to such a collection of $2j$ stars written in the spherical coordinates $\{\theta_k,\phi_k \}_{k=1}^{2j}$ as the Majorana representation of a state $\ket{\psi}$, and denote it by
\begin{equation}
    \mathcal{M} (\ket{\psi} ):=\{\theta_k,\phi_k \}_{k=1}^{2j}.\label{MajoranaRepresentation}
\end{equation}
For instance, the state $\ket{j,j}$ corresponds to the trivial Majorana polynomial $w(z)=z^{2j}$, which has $2j$-degenerated root at $z=0$, and hence is represented by $2j$ stars at the north pole. More generally, the Majorana polynomial corresponding to $\ket{j,m}$ state has two degenerated roots at $z=0$ and $z=\infty$ with multiplicities $j+m$ and $j-m$ respectively. Hence the state $\ket{j,m}$ is represented by $j+m$ stars at the north and $j-m$ stars at the south pole. 
Note that for a spin-$\frac{1}{2}$ particle, the Majorana representation, reduces to the Bloch representation of a state provided in Eq.~\cref{eq1}.

\begin{figure}[h]
  \includegraphics[width=1\columnwidth]{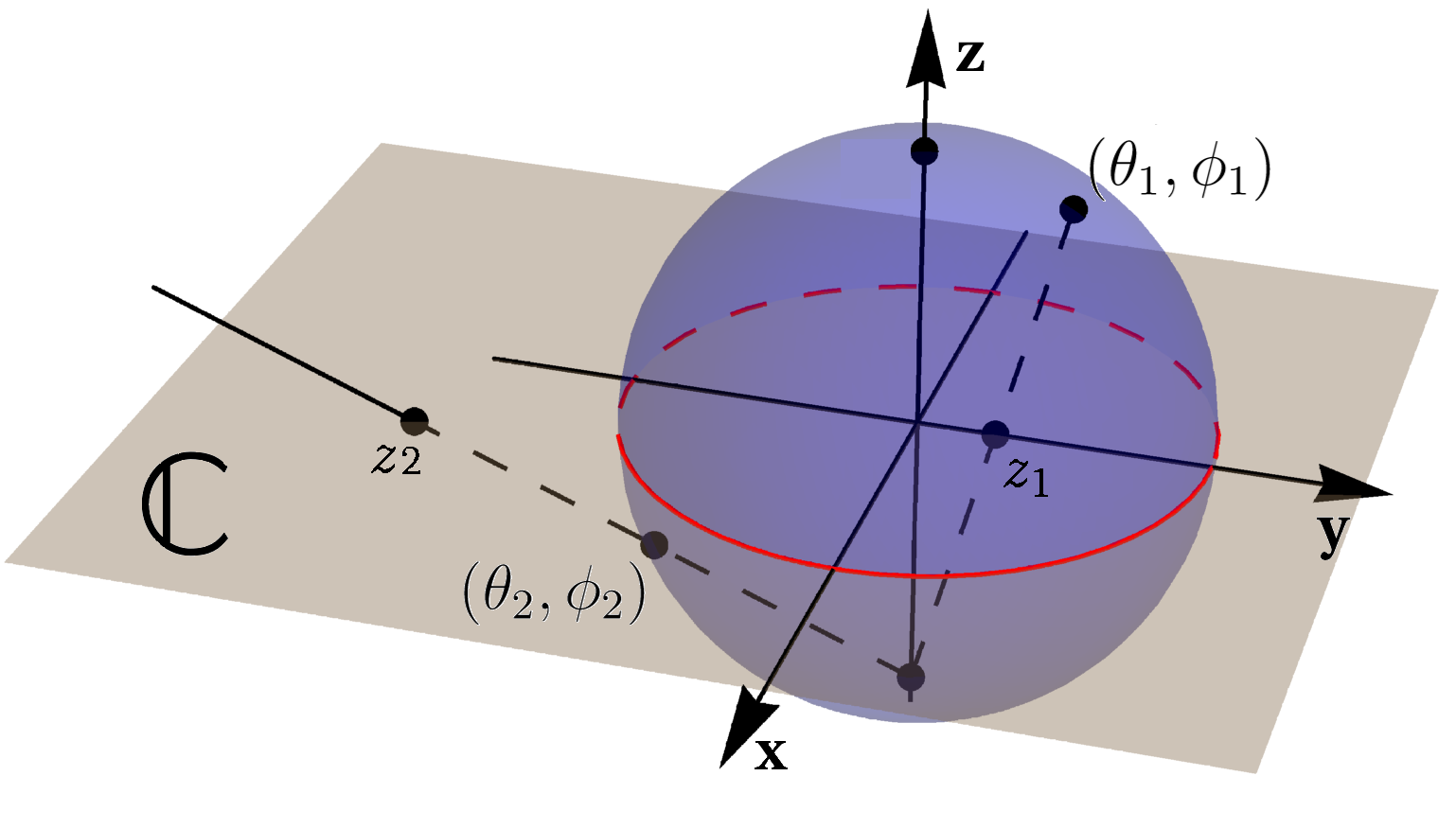}
  \caption{A sphere with the stereographic projection of points $z_1=e^{i\phi_1}\tan(\theta_1/2)$ and $z_2=e^{i\phi_2}\tan(\theta_2/2)$ presented.}
  \label{stereo}
\end{figure}

\subsection{The stellar representation for symmetric multi-qubit states}

The stellar representation has a natural interpretation while identifying a spin-$j$ system with a symmetric subspace of a system of $2j$ qubits. 
The symmetric subspace of dimension $2j+1$ is spanned by \textit{the Dicke states} $\ket{D_{2j,k}}$, which are a uniform superposition of states with a given number of $k$ excitations \cite{Dicke},
\begin{equation}
   \ket{D_{2j,k}}=\binom{2j}{k}^{-\frac{1}{2}}\sum_{\sigma \in \mathbb{S}_{2j}}\sigma\bigl(\ket{0}^{\otimes 2j-k}\otimes\ket{1}^{\otimes k}\bigr).
\end{equation}
Here $\sigma\in \mathbb{S}_{2j} $ denotes a permutation of subsystems determined by an element of the symmetric group $\mathbb{S}_{2j}$. Any symmetric state may be uniquely expressed as a combination of Dicke states,
\begin{equation}
\label{SymState}
    \ket{\psi_{sym}}=\sum_{k=0}^{2j}Z_{k}\ket{D_{2j,k}}
    \;\in \;\mathcal{H}_2^{\otimes 2j},
\end{equation}
with normalization $\sum_{k=0}^{2j}|Z_{k}|^2=1$. 
By identifying the eigenstates of the $J_z$ operator with Dicke states
\begin{equation}
\label{Identification}
F\;:\; \ket{j,m} \;\mapsto \; \ket{D_{2j, j-m}},
\end{equation}
one can relate a $\mathcal{H}_{2j+1}$ state space with symmetric subspace of $ \mathcal{H}_{2}^{\otimes 2j} $, and hence a state $\ket{\psi}\in\mathcal{H}_{2j+1}$ with a symmetric state $\ket{\psi_{sym}}$ of $2j$ qubits, --~see Eqs.~\cref{state-ang-mom-bas,SymState}. Above equation determines the isomorphism $F$ of two spaces, and for any spin-$j$ state $\ket{\psi}$ we denote by $\ket{F( \psi )} $ the related symmetric state of $2j$ qubits. As we shall see, both states have the same \textit{Majorana representation}, $\mathcal{M} (\ket{\psi} )=\mathcal{M} (\ket{F(\psi )} )$. 

Interestingly, there is another way of presenting a symmetric state of $2j$ qubits. Consider the collection of $2j$ quantum states
\begin{equation}
\label{SymThree}
    \ket{\Phi_{k}}=\cos(\tfrac{\theta_k}{2})\ket{0}+e^{i\phi_k}\sin(\tfrac{\theta_k}{2})\ket{1},
\end{equation}
with $\theta_k \in [0,2\pi ], \phi_k \in [0,\pi ]$ and $k=1,\ldots, 2j$. A symmetric superposition of their tensor products constitutes a symmetric state
\begin{equation}
\label{SYmTwo}
    \ket{\psi_{sym}}=\mathcal{N}\sum_{\sigma \in \mathbb{S}_{2j}}\ket{\Phi_{\sigma(1)}}\otimes\cdots\otimes\ket{\Phi_{\sigma(2j)}},
\end{equation}
where $\sigma \in \mathbb{S}_{2j} $ runs over all permutations of indices, and $\mathcal{N}$ denotes the normalization factor. A connection between those two representations of symmetric states (\ref{SymState}) and (\ref{SYmTwo}) is given by the Majorana representation. Consider a symmetric state (\ref{SymState}) with state coefficients $Z_k$. The related Majorana polynomial $w(z)$, Eq.~\cref{polynomial}, has $2j$ roots $z_i$ with respect to $z$ variable. On one hand, the stereographic projection $z_k \mapsto (\theta_k ,\phi_k )$, given by Eq.~\cref{stereographicProj}, maps roots of the polynomial $w(z)$ onto the surface of a sphere. On the other hand, however, the set of angles $(\theta_k ,\phi_k )$ provides an alternative description of the symmetric state, as symmetrization  (\ref{SYmTwo}) of $2j$ qubit states (\ref{SymThree}), determined by the angles $(\theta_k ,\phi_k )$.

\section{Measures of quantumness}
\label{Anticoherence_measures}

In this section, we introduce the concept of anticoherence in spin-$j$ system and provide its quantitative description. Furthermore, we extended those concepts to the notion of anticoherence on an orthonormal basis to quantify its quantumness. 

A spin coherent state $\ket{\psi}\in\mathcal{H}_{2j+1}$ that points in direction $\mathbf{n}$ in $\mathbb{R}^3$ is a state $\ket{\mathbf{n}}$ for which the polarization vector $\mathbf{p}$ is of length $j$ i.e. 
\begin{equation}
    \mathbf{p}\equiv \braket{\mathbf{n}|\mathbf{J}|\mathbf{n}} = j\mathbf{n},
\end{equation} 
where $\mathbf{J}=(J_x,J_y,J_z)$ is the spin-$j$ operator, and $\hbar$ is set to unity. In the Majorana representation, a spin-$j$ coherent state is represented by $2j$ degenerated stars on a sphere. Their position is given by the vector $\mathbf{n}$. 
A spin state $\ket{\psi}$ is anticoherent, if its polarization vector vanishes, $\mathbf{p}=\mathbf{0}$. One may introduce higher orders of anticoherence, namely a spin state $\ket{\psi}$ is called \textit{t-anticoherent} \cite{Zimba} if $\braket{\psi|(\mathbf{n}\cdot\mathbf{J})^k|\psi}$ is independent of $\mathbf{n}$ for $k=1,\ldots,t$.

For a pure symmetric quantum state of $2j$ subsystems $\ket{\psi}\in \mathcal{H}_{2}^{\otimes 2j}$, we consider a t-partite reduced density operator,
\begin{equation}
    \rho_t (\psi ):= \text{tr}_{1,\ldots , 2j-t} \,  \ket{\psi} \bra{\psi},
\end{equation}
and analyze its purity, 
\begin{equation}
\label{EqAuxi}
R_t(\ket{\psi}) := \text{tr}\, \big(\rho_t (\psi )^2 \big).
\end{equation} 
This quantity can be used to quantify the coherence of the related system of spin-$j$ particle \cite{GiraudTensor}. Recall that the isomorphism \cref{Identification} identifies a spin-$j$ system with a symmetric state of $2j$ qubits. Thus, Baguette and Martin \cite{Baguette-measure} introduced the following measures of anticoherence of order $t\geq 1$ based on the purity of the reduced state:
\begin{equation}
\label{anticoherence}
    \mathcal{A}_t(\ket{\psi})=\frac{t+1}{t}[1-R_t(F(\ket{\psi}))].
\end{equation}
where $\ket{\psi}\in \mathcal{H}_{2j+1} $ is a spin-$j$ system, and \\$F (\ket{\psi})\in \mathcal{H}_{2}^{\otimes 2j} $ denotes the corresponding symmetric state, --~see the map \cref{Identification}.
As discussed in Ref. \cite{Baguette-measure} this quantity enjoys the following properties,
\begin{enumerate}
\item $\mathcal{A}_t(\ket{\psi})=0\iff \ket{\psi}$ is coherent,
\item $\mathcal{A}_t(\ket{\psi})=1\iff \ket{\psi}$ is t-anticoherent,
\item $\mathcal{A}_t(\ket{\psi})\in[0,1]$ for all $\ket{\psi}$.
\item $\mathcal{A}_t(\ket{\psi})$ is invariant under phase changes and spin rotations,
\end{enumerate}
and hence provides a plausible measure of $t$-anticoherence. 
Making use of this quantity we propose the quantumness measure $\mathcal{B}_t$ for an orthonormal basis using the arithmetic mean of $t$-anticoherence measure of constituting states
\begin{equation}
    \mathcal{B}_t(U)=\sum_{i=1}^{N} \frac{\mathcal{A}_t(\ket{\psi_i})}{N}.
    \label{anticoherence-measure}
\end{equation}
Here $N\text{=}2j+1$ denotes the dimension of the Hilbert space,  $U$ is a unitary matrix that represents a basis and $\ket{\psi_i}$ is i-th state in this basis. Observe that the higher value of the quantity $ \mathcal{B}_t(U)$, defined in Eq.~\cref{anticoherence-measure}, the more "quantum" (anticoherent), is the analyzed basis determined by the unitary matrix $U$. It is easy to note that if a basis represented by $U$ gives $\mathcal{B}_k(U)=1$, then all the vectors forming the basis are $k-$anticoherent.

\section{Bases of extreme quantumness}
\label{Main_results}

In this section, we present the most "classical" bases characterized by the smallest values of quantumness. Furthermore, we identify also the bases of the maximal quantumness. Note that such bases can be interpreted as orthogonal measurements of extreme quantumness. We present both, numerical and analytical results in dimensions $N=3,4,5,6,7$, which correspond to spin $j=1,\tfrac{3}{2},2,\tfrac{5}{2},3$. For convenience, we present the basis vectors $\{\ket{\psi_i} \}$ in the form of a unitary matrix $U$, where the i-th column of the matrix $U$ corresponds to the i-th vector $\ket{\psi_i}$ in the basis expressed in the angular momentum basis $J_z$, i.e. $U=(U_{ki})$, where $U_{ki}=\braket{j,m|\psi_i}$, with $m=j+1-k$. Unitary matrices corresponding to the basis maximizing (minimizing) the quantumness $\mathcal{B}_1$ in dimension $N=2j+1$ shall be denoted by $U_N^{q}$ and $U_N^{c}$ respectively. In general, bases of extreme values of a given measure of quantumness correspond to highly symmetric constellations of stars in the Majorana representation, --~see Fig. (2-10).

Note that the measure of quantumness are invariant under $SU(2)$ transformations represented as an action of the D-matrices of Wigner. Any such action corresponds to the rotation of a sphere in the Majorana representation, for more details see \cref{Appendix E,Appendix F}.

Starting from $N=4$, we use a numerical algorithm inspired by a random walk to find bases of extreme quantumness with respect to the measure of 1-anticoherence $\mathcal{B}_1$. To find a basis that maximizes quantumness we choose a random unitary matrix $U_0$, that represents an orthonormal basis, then we make a random step
\begin{equation}
   U_0\rightarrow U_1=U_0\exp{(iHa)},
\end{equation}
where $H$ is a random hermitian matrix from Gaussian Unitary Ensemble, while a real parameter $a$ is a small time step. If $\mathcal{B}_1(U_1)>\mathcal{B}_1(U_0)$, then we treat $U_1$ as new basis $U_0$ and repeat the procedure. Otherwise, we pick another random hermitian matrix $H$ and repeat the above steps. During the procedure, the parameter $a$ is decreased in a way to obtain an extremal basis with increased precision. Usually, the number $a$ varied from $0.1$ to $10^{-15}$. Analogously, one can obtain bases that minimize the average coherence using a similar approach. A simplified version of our code is given below.

\noindent\rule{8.2cm}{0.4pt}
\begin{lstlisting}[language=Python]
#basis
U0
#number of steps
n1
n2
#step length
a
Do[
    Do[
        H=RandomVariete[GUE[N]]
        U1=U0exp[i*H*a]
        If[B[U1]>B[U0], U0=U1]
        ,n1]
    a=a/2
    ,n2]
\end{lstlisting}
\noindent\rule{8.2cm}{0.4pt}
To show that a final basis $U_0$ yields a local extremum, one should check that the gradient vanishes, $\nabla \mathcal{B}_1(U_0)=\mathbf{0}$ and the Hessian is negative definite (for maximum) or positive definite (for minimum). For further details see \cref{Appendix D}.

It turns out that for higher dimensions, starting from $N=5$, the measure of quantumness $\mathcal{B}_1$ is saturated by many different bases without any visible symmetries. This implies that several bases can be formed by 1-anticoherent states. Therefore, it is natural to ask whether it is possible to obtain a maximal value of $\mathcal{B}_2$, keeping $\mathcal{B}_1=1$. Interestingly, by requiring both measures $\mathcal{B}_1$ and $\mathcal{B}_2$ to increase in a single step of the algorithm we managed (except for the case $N=6$) to obtain bases consisting states of higher orders of anticoherence with interesting symmetries.

\subsection{Extremal bases for N=3}
\begin{figure}[h]
  \centering
  \includegraphics [width=1\columnwidth] {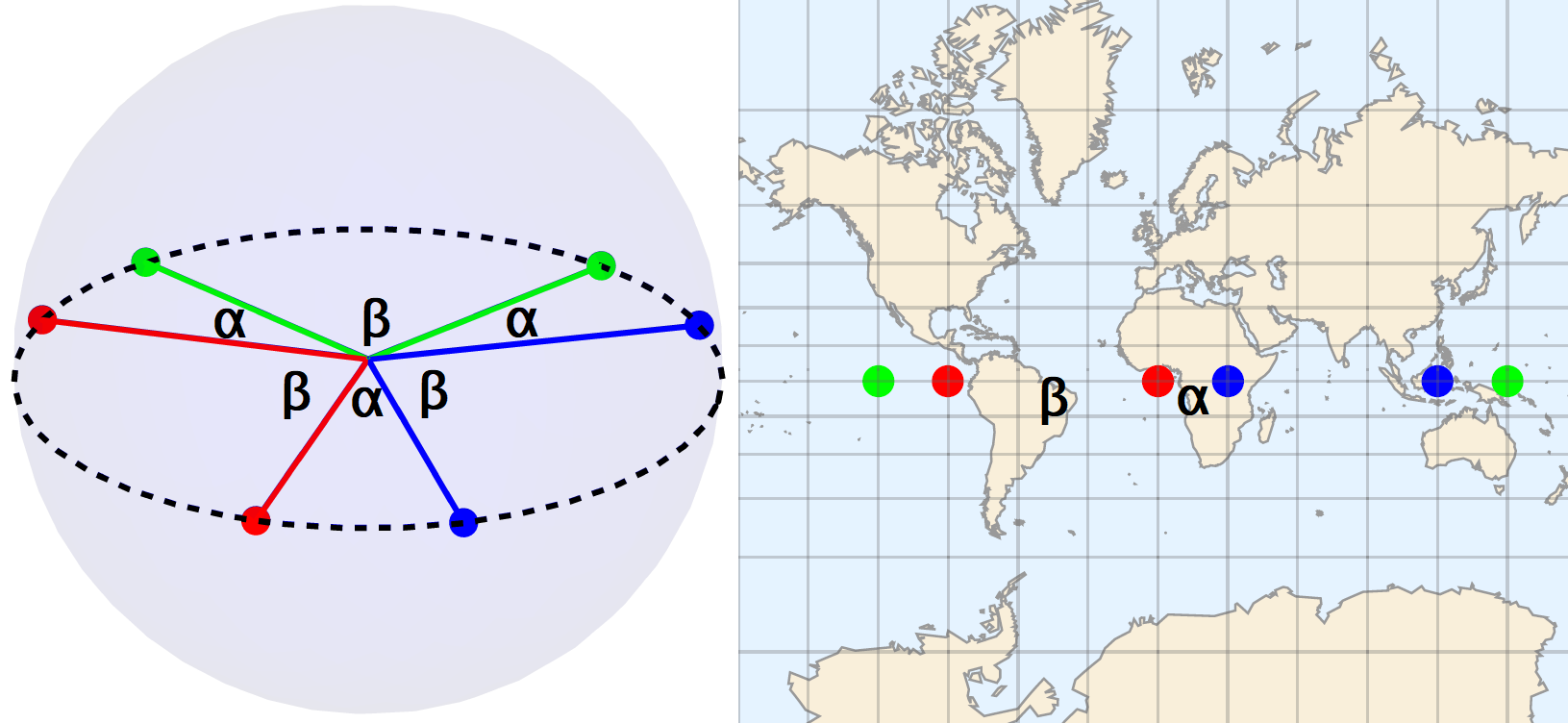}
  \caption{\small The most classical basis \cref{U-3-max} in $\mathcal{H}_3$ $(j=1)$ is represented by three pairs of stars in the Majorana representation. Each state is presented by two dots of a given color at the equator. The angles between lines connecting stars with the center of the sphere read $\alpha\text{=}\pi/6$ and $\beta\text{=}\pi/2$. The left panel shows the stars on the sphere while the right one uses the Mercator projection, with the geographic grid drawn.}
  \label{j=1-coh-basis}
\end{figure}

Consider an orthonormal basis in $\mathcal{H}_3$, corresponding to $j=1$. Up to $SU(2)$ transformations, corresponding to the rigid rotation of the entire sphere, any basis can be parametrized by three real parameters, --~see \cref{Appendix A}.
The $\mathcal{B}_t$ measure is invariant under the $SU(2)$ transformations, so one may use this parametrization to find bases of extreme quantumness. 
Therefore, the problem reduces to finding the global extrema of a function of three real variables $\mathcal{B}_1(\Theta_1,\Theta_2,\Phi)$, which we solved analytically. 
The most classical basis, for which $\mathcal{B}_1=\frac{1}{9}$, consists the following states,
\begin{align}
\nonumber
       \ket{\psi^{c}_3}=\frac{1}{\sqrt{3}}(\ket{1,1}+\ket{1,0}+\ket{1,\text{-}1}),~~~~~~~~~\hspace{0.5mm}
\\
\label{U-3-max-stany}
       \ket{\psi^{\prime ~c}_{~3}}=\frac{1}{\sqrt{3}}(\ket{1,1}+\omega_3\ket{1,0}+\omega_3^2\ket{1,\text{-}1}),\hspace{0.8mm}
\\
\nonumber       
       \ket{\psi^{\prime\prime ~c}_{~3}}=\frac{1}{\sqrt{3}}(\ket{1,1}+\omega_3^2\ket{1,0}+\omega_3\ket{1,\text{-}1}),
    \end{align}
where $\omega_3=e^{2\pi i/3}$. Basis represented in the Majorana representation is depicted in \cref{j=1-coh-basis}. Each state in this basis can generate another two states by rotation around $\hat{z}$ axis by angle $2\pi/3$ and $4\pi/3$. All stars lie on the equator. This basis corresponds to the \textit{Fourier} matrix,
\begin{align}
    U_3^{c}=F_3= \frac{1}{\sqrt{3}}\left(\begin{matrix} 1&1&1\\1&\omega_3&\omega_3^2\\1&\omega_3^2&\omega_3\end{matrix}\right).
      \label{U-3-max}
  \end{align}
 
\begin{figure}[h]
  \centering
  \includegraphics [width=1\columnwidth] {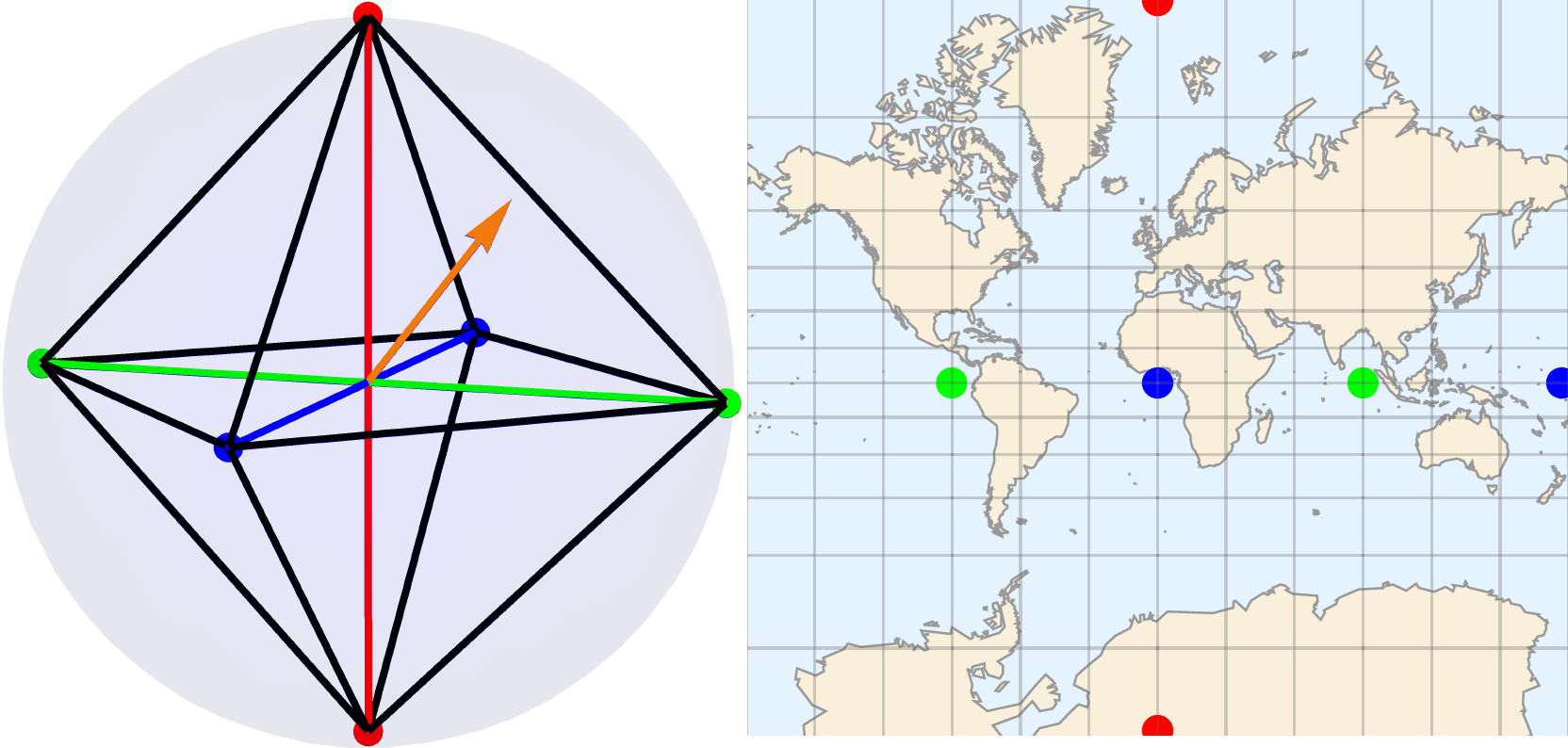}
  \caption{\small The most quantum basis \cref{U-3-min} in $\mathcal{H}_3$ $(j=1)$ represented by three pairs of points on the sphere (left). The same configuration in the Mercator projection is shown on the right. The Orange arrow represents one of the rotational axes that may be used to generate the entire basis by rotation of one state by $2\pi/3$ and $4\pi/3$.}
  \label{j=1-ant-basis}
\end{figure}

The most quantum basis, for which the measure $\mathcal{B}_1=1$, reads
\begin{align}
\nonumber
       \ket{\psi^{q}_3}=\ket{1,0},~~~~~~~~~~~~~~~~\hspace{4mm}
\\
\label{U-3-min-stany}
      \ket{\psi^{\prime ~q}_{~3}}=\frac{1}{\sqrt{2}}(\ket{1,1}+\ket{1,\text{-}1}),\hspace{0.85mm}
\\
\nonumber       
      \ket{\psi^{\prime\prime ~q}_{~3}}=\frac{1}{\sqrt{2}}(\ket{1,1}-\ket{1,\text{-}1}).
    \end{align}
This basis corresponds to the following unitary matrix
\begin{equation}
      U_3^{q}=\frac{1}{\sqrt{2}}\left(
      \begin{matrix} 0&1&1\\
      \sqrt{2}&0&0\\
      0&1&-1
      \end{matrix}\right),
  \label{U-3-min}
\end{equation}
In the stellar representation, the points representing a single state lie on a line, going through the center of a sphere that graphically corresponds to 1-simplex $\Delta_1$. The entire constellation spans a regular octahedron, --~see \cref{j=1-ant-basis}. This basis may be generated by rotating one of its elements around a vector directed to the center of any face of a regular octahedron by $2\pi/3$. Observe that the same constellation represents 3 mutually unbiased bases (MUB) in $\mathcal{H}_2$ i.e. two points of the same color correspond to a single basis.

Since the least quantum basis in dimension $N=3$ was found as an analytical solution of minimizing the $\mathcal{B}_1(\Theta_1,\Theta_2,\Phi)$ function, while the most quantum basis saturates the bound for the measure $\mathcal{B}_1$, we can conclude this reasoning with the following statement.

\begin{proposition}
The least quantum and the most quantum bases in $\mathcal{H}_3$ are presented in Eqs.~\cref{U-3-max-stany} and \cref{U-3-min-stany}, and correspond to $U_3^{c}$ and $U_3^{q}$ respectively. The measure of coherence $\mathcal{B}_1$ archives values $\mathcal{B}_1 (U_3^{c})=\frac{1}{9}$ and $\mathcal{B}_1(U_3^{q})=1$.
\end{proposition}

I a close analogy to the notion of iso-entangled bases for the multiqubit systems \cite{Iran,Grzegorz,Kuba,SCZG23} we will introduce a related notion. An orthogonal basis $\{\ket{\psi_i}\}$ in $\mathcal{H}_{N}$, determined by unitary $U\in U(N)$ is called \textit{iso-coherent,} if all its vectors $\ket{\psi_i}$ are equivalent with respect to the $SU(2)$ group, so that for any measure of spin-coherence all its values are equal.

\subsection{Extremal bases for N=4}

 \begin{figure}[h]
  \centering
  \includegraphics [width=1\columnwidth] {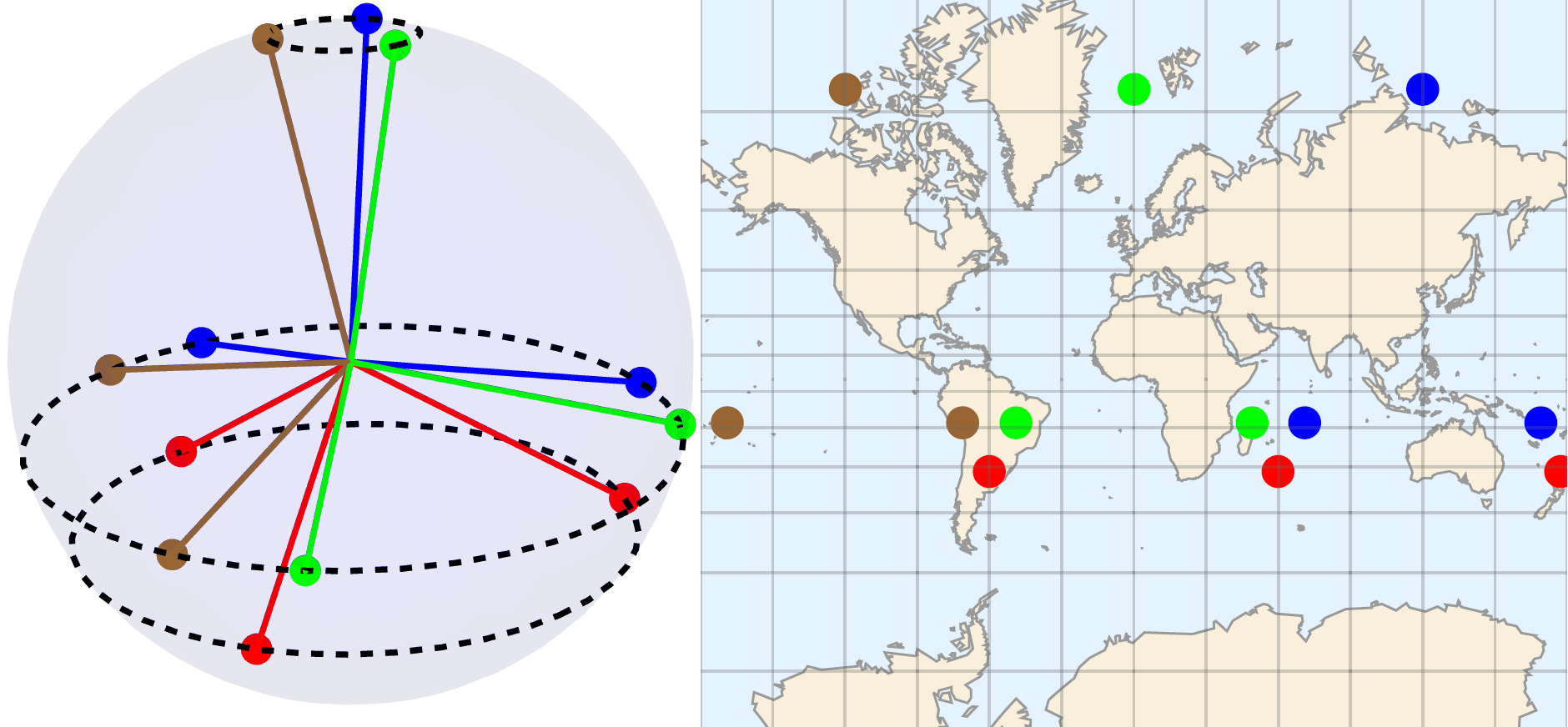}
  \caption{\small  The most classical basis \cref{U-4-max} in $\mathcal{H}_4$ $(j=3/2)$ represented by 4 triples of points. The sphere (left) and the Mercator projection (right). }
  \label{Dim-4-snopy}
\end{figure}

According to the quantum walk algorithm, the most classical basis for $N=4$ ($j=3/2$) was found, for which $\mathcal{B}_1=\frac{1}{9}$, --~see \cref{Dim-4-snopy}. This basis is formed by four states which are equivalent up to $SO(3)$ rotation on a sphere.  This basis corresponds to the following unitary matrix
\begin{equation}
   \begin{aligned}
      U_4^{c}\text{=}\frac{1}{\sqrt{18(3\text{-}2\sqrt{2})}}\left(\begin{matrix} 1&1&1&\tau\sqrt{3}\\\nu&\nu \omega_3&\nu \omega^{2}_3&0\\\nu&\nu \omega^{2}_3&\nu \omega_3&0\\\tau&\tau&\tau&\text{-}\sqrt{3}\end{matrix}\right),
  \end{aligned}
  \label{U-4-max}
\end{equation}
where $\nu=\sqrt{3}(2-\sqrt{2})$, $\tau=3-2\sqrt{2}$ and $\omega_3=e^{i2\pi/3}$.
This result was obtained by an observation of specific symmetries of purely numerical expression and then using them as constraints to reduce the problem to finding an extremum of a one parameter function, which allowed us to obtain an analytical solution for the optimal basis.

\begin{figure}[h]
  \includegraphics [width=1.0\columnwidth] {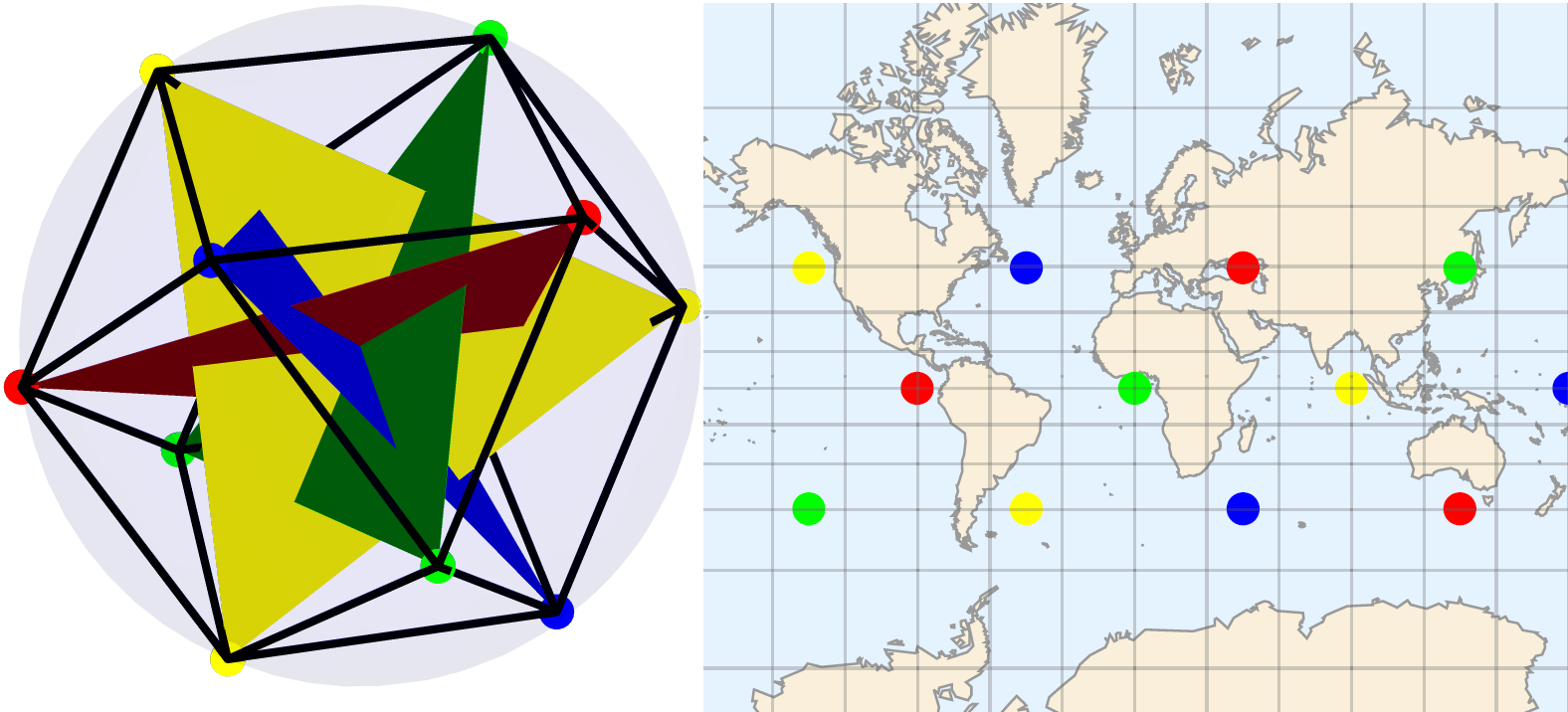}
  \caption{\small The most quantum basis in  $\mathcal{H}_4$ $(j=3/2)$ represented by $4\times3=12$ points. The sphere (left) and the Mercator projection (right).  Three stars representing a single state span an equilateral triangle. Constellation of the entire basis spans cuboctahedron, the edges are denoted by black lines.}
  \label{Dim-4-anticoherent}
\end{figure}

The most quantum basis, --~see \cref{Dim-4-anticoherent}, for which \\$\mathcal{B}_1=1$, corresponds to following unitary matrix
\begin{equation}
   \begin{aligned}
      U_4^{q}=\frac{1}{\sqrt{6}}\left(\begin{matrix} \sqrt{3}&1&1&1\\0&\xi&\xi\omega_3&\xi\omega_3^2\\0&\xi&\xi\omega_3^2&\xi\omega_3\\\text{-}\sqrt{3}&1&1&1\end{matrix}\right),
  \end{aligned}
  \label{U-4-min}
\end{equation}
where $\omega_3=e^{i2\pi/3}$ and $\theta_3=\arcsin{(1/\sqrt{3})}$ with \\$z_3=\tan\frac{\theta_3}{2}e^{-i5\pi/6}$ and $\xi=(1+z_3+1/z_3)/\sqrt{3}$.

In analogy to the most classical basis, the four states are equivalent up to $SO(3)$ rotation on a sphere. The entire basis may be generated by rotating a single state by multiples of $\pi/2$ around the $\mathbf{z}$ axis. Each vector of the basis is represented by three stars that form an equilateral triangle, then graphically it corresponds to the 2-simplex $\Delta_2$. The entire constellation of 12 stars forms an Archimedean solid, called cuboctahedron, --~see \cref{Dim-4-anticoherent}. Therefore this basis belongs to the class of iso-coherent bases.

The most quantum basis $U_4^{q}$ saturates the bound for the measure $\mathcal{B}_1$, which leads to the following statement. 
\begin{proposition}
The most quantum basis in $\mathcal{H}_4$ is given by Eq.~\cref{U-4-min} for which $\mathcal{B}_1=1$.  
\end{proposition}
\begin{conjecture}
The matrix $U_4^{c}$ given in \cref{U-4-max}, for which  $\mathcal{B}_1=\frac{1}{9}$, leads to the most spin-coherent (classical) basis in $\mathcal{H}_4$. 
\end{conjecture}
\subsection{Extremal bases for N=5}
\begin{figure}[h]
  \centering
  \includegraphics [width=1\columnwidth] {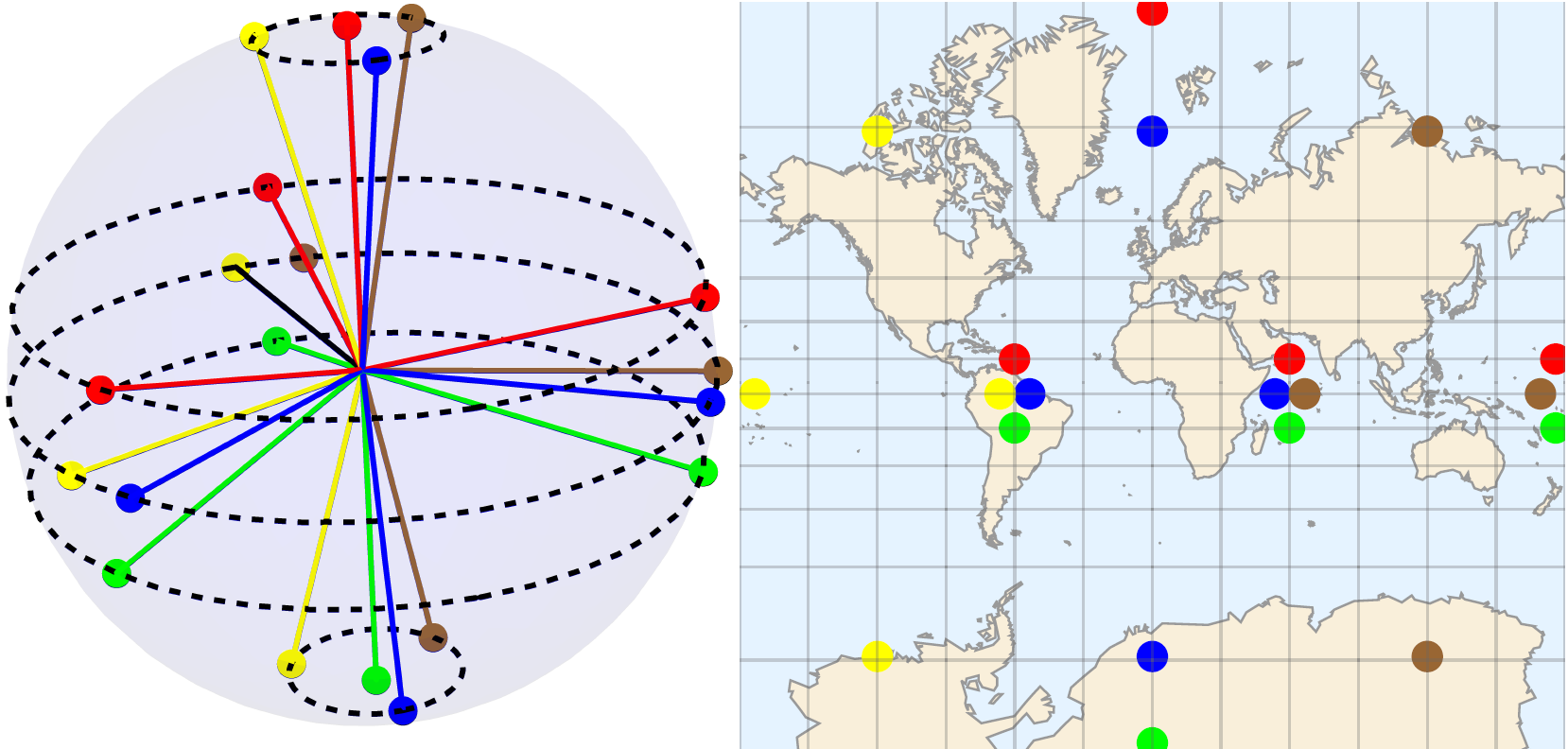}
  \caption{\small The most classical basis in  $\mathcal{H}_5$ $(j=2)$ represented by $5\times4=20$ points. The sphere (left) and the Mercator projection (right).}
  \label{Dim-5-snopy}
\end{figure}

Using the numerical procedure described above we found the most classical basis presented in \cref{Dim-5-snopy}. The set of states forming this basis can be divided into two equivalence classes, with respect to rotation of the sphere. The first class contains two states
\begin{equation}\label{Dim-5-max-stan1}
   \ket{\psi^{c}_5}=\mathcal{N}_1 (\ket{2,2}+\frac{r_1^3}{2}\ket{2,\text{-}1}),
\end{equation}
where $r_1\in\mathbb{R}$ is the only possible parameter that does not change the symmetry of the state and $\mathcal{N}_1$ is the normalization constant. The other state is obtained by rotation around the $\mathbf{x}$ axis by angle $\pi$. The second class contains the state 
\begin{align}\label{Dim-5-max-stan2}
\ket{\psi^{\prime ~c}_{~5}}=\mathcal{N}_2 (\ket{2,2}+\chi\ket{2,1}+\nonumber\\+\frac{\chi'}{\sqrt{6}}\ket{2,0}
+\chi\ket{2,\text{-}1}+\ket{2,\text{-}2})
    \end{align}
and its rotations by angles $2\pi/3$ and $4\pi/3$ around the $\mathbf{z}$ axis. Here, $\mathcal{N}_2$ is the normalization constant, $\chi=(r_2+1/r_2+2\cos\phi_3)/2$, \\$\chi'=2((r_2+1/r_2)\cos\phi_3+1)$, and $r_2,\phi_3\in\mathbb{R}$ are parameters of the symmetry of this class of states. 

In the stellar representation, the state $\ket{\psi^{ c}_{5}}$ corresponds to one star at the north pole, and another three equally distributed at a circle of latitude $\theta_4=\arctan(2r_1)$, where $r_1$ is determined numerically --~see \cref{Appendix B} and red stars in \cref{Dim-5-snopy}. Similarly $\ket{\psi^{\prime ~q}_{~5}}$ is represented by one star which horizontal angle $\theta_5=\arctan(2r_2)$. Horizontal angle of the second star reads $\theta_6=\pi-\theta_5$. The remaining two stars lie on the equator with azimuthal angles $\phi_3$ and $\text{-}\phi_3$, --~see blue stars in \cref{Dim-5-snopy}.

Imposing these symmetries and orthogonality conditions reduces the number of degrees of freedom to a single one. The minimum of the quantumness measure obtained numerically reads, $\mathcal{B}_1\approx0.126$ --~see \cref{Appendix B}. Anticoherence measure values for individual states of the basis are $\mathcal{A}_1(\ket{\psi^{c}_5})\approx0.139$ and $\mathcal{A}_1(\ket{\psi'^{c}_5})\approx0.118$. Numerical results allow us to advance the following statement.
\begin{conjecture}
The  most classical basis in $\mathcal{H}_5$ is $U_5^{c}$, for which $\mathcal{B}_1\approx0.126$.
\end{conjecture}

\begin{figure}[h]
  \centering
  \includegraphics [width=1\columnwidth] {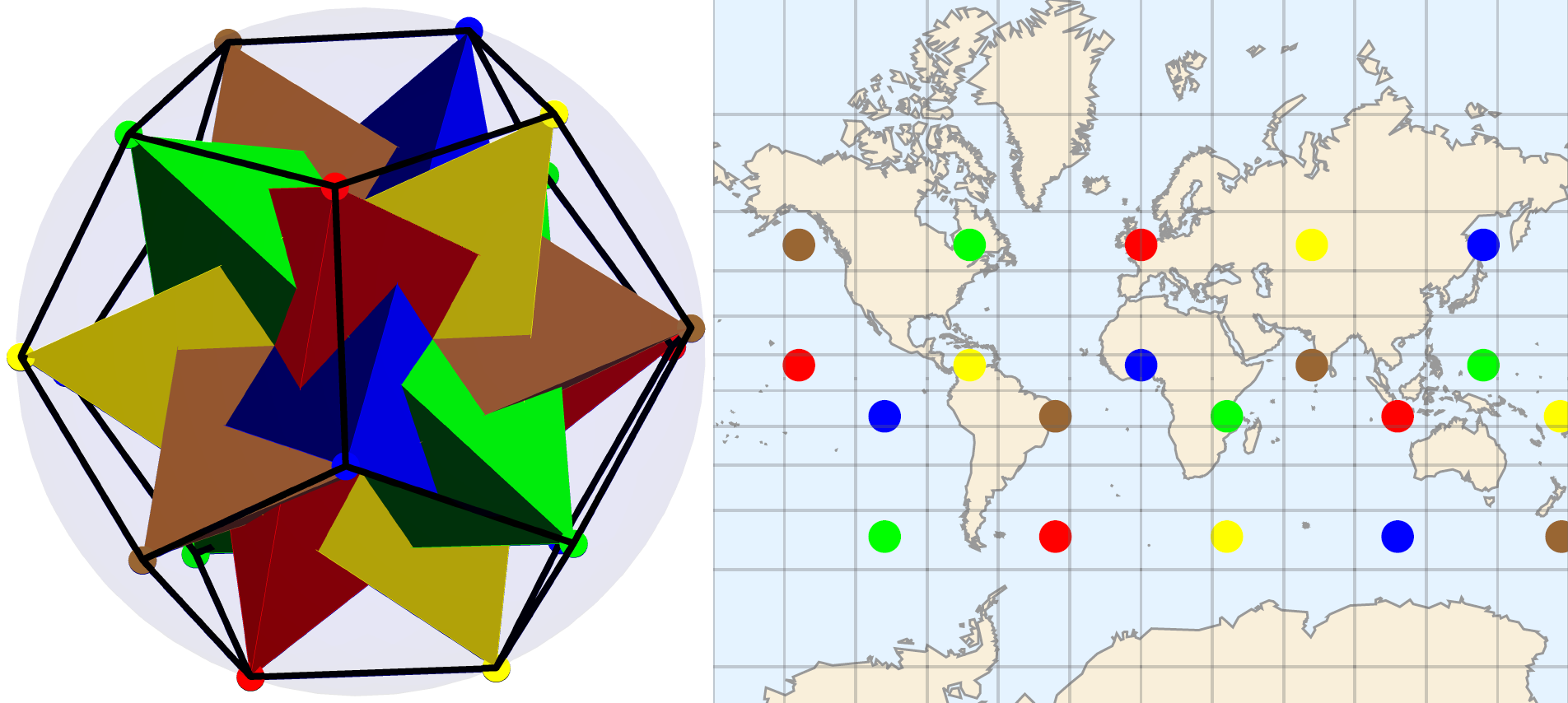}
  \caption{\small The most quantum basis \cref{U-5-min} in  $\mathcal{H}_5$ $(j=2)$ represented by $5\times4=20$ points on the sphere (left) and in the Mercator projection (right). Stars representing each state span a regular tetrahedron and the entire basis forms a compound of five tetrahedra and spans a regular dodecahedron, plotted with solid lines.}
  \label{Dim-5-dodecahedron}
\end{figure}

Numerical analysis provided us several different bases saturating measure of quantumness (1-anticoherence) $\mathcal{B}_1=1$. After imposing condition to increase both measures $\mathcal{B}_1$ and $\mathcal{B}_2$ in a single algorithm step we arrived at the basis for which $\mathcal{B}_2=\mathcal{B}_1=1$. Exactly the same basis was earlier described by Zimba in \cite{Zimba}. It is formed by five states equivalent up to rotation on the sphere. Each of those states is represented by four stars that form a regular tetrahedron, then graphically it corresponds to 3-simplex $\Delta_3$. It may be constructed by rotation of a state
\begin{equation}
   \ket{\psi^{tet}_5}=\frac{1}{\sqrt{3}} (\ket{2,2}+\sqrt{2}\ket{2,\text{-}1}).
\end{equation}
The corresponding unitary matrix reads
\begin{equation}
   \begin{aligned}
       U_5^{q}\text{=}\frac{1}{\sqrt{5}}\left(\begin{matrix} 1&1&1&1&1\\\text{-}\kappa&\text{-}\kappa\omega_5&\text{-}\kappa\omega_5^2&\text{-}\kappa\omega_5^3&\text{-}\kappa\omega_5^4\\\lambda&\lambda\omega_5^2&\lambda\omega_5^4 &\lambda\omega_5^6 &\lambda\omega_5^3\\\kappa&\kappa\omega_5^3&\kappa\omega_5^6&\kappa\omega_5^4&\kappa\omega_5^2\\1&\omega_5^4&\omega_5^3&\omega_5^2&\omega_5\end{matrix}\right),
  \end{aligned}
  \label{U-5-min}
\end{equation}
where $\kappa=\frac{1}{4}(1+i\sqrt{15})$, $\lambda=\sqrt{\text{-}\kappa}$ and $\omega_5=e^{i2\pi/5}$.
The entire constellation of basis consists of 20 stars and forms a regular dodecahedron, --~see \cref{Dim-5-dodecahedron}. Above reasoning leads to the following statement.

\begin{proposition}
    The most quantum basis in $\mathcal{H}_5$ is $U_5^{q}$, for which  $\mathcal{B}_1=\mathcal{B}_2=1$.
\end{proposition}
Note that a regular dodecahedron arises in quantum theory in several contexts, including the Penrose dodecahedron \cite{Penrose,Penrose1,Massad} formed by 40 pure states in $\mathcal{H}_4$ which allow one to construct a proof of the Bell's nonlocality theorem. The same configuration is related to the set of five iso-entangled two-qubit mutually unbiased bases, as the partial traces of 20 pure states in $\mathcal{H}_4$ lead to a regular  
dodecahedron inside the Bloch ball \cite{Kuba}.

\subsection{Extremal bases for N=6}
\begin{figure}[h]
  \centering
  \includegraphics [width=1\columnwidth] {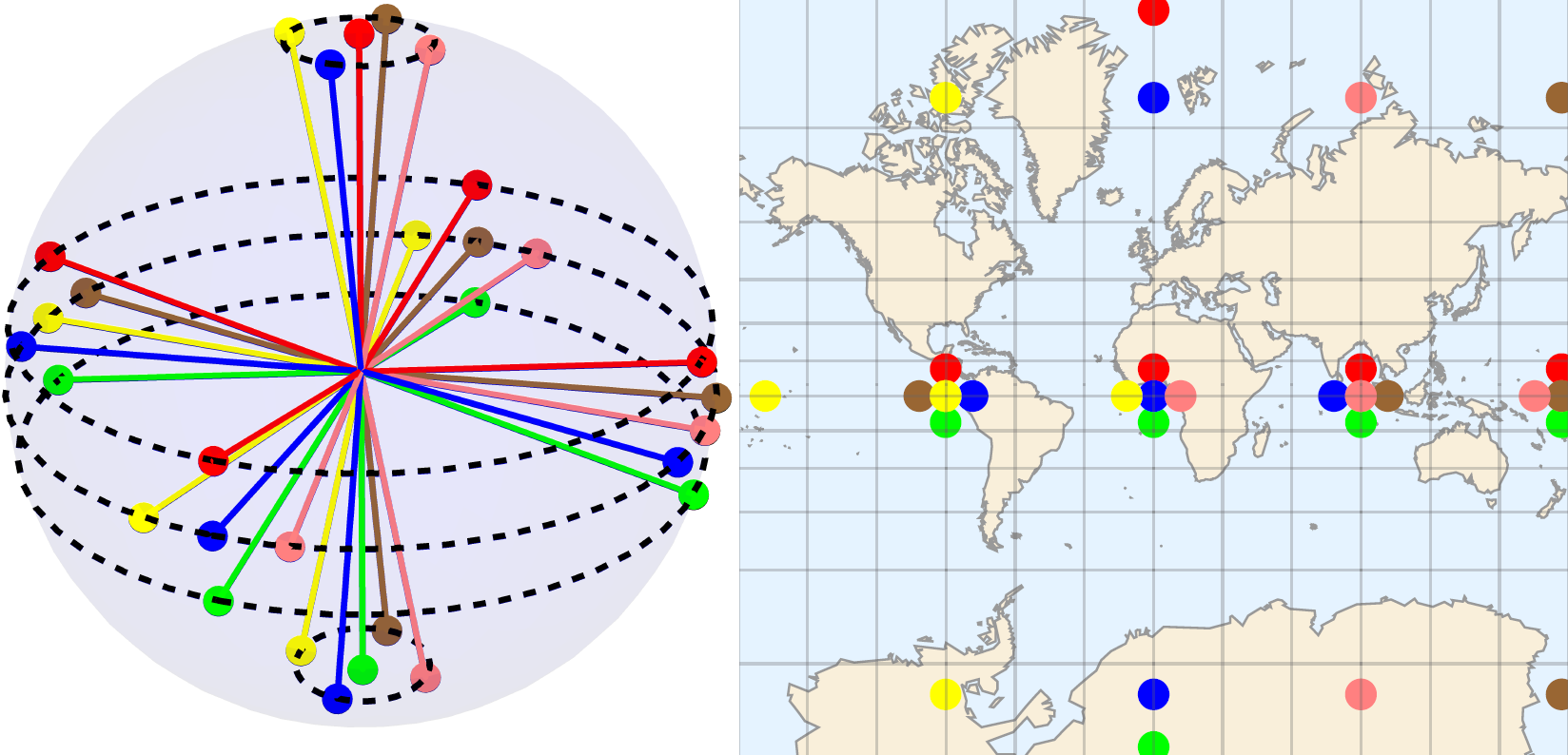}
  \caption{\small A  most classical basis in  $\mathcal{H}_6$ $(j=5/2)$ represented by $6\times5=30$ points. The sphere (left) and the Mercator projection (right).}
  \label{Dim-6-snopy}
\end{figure}

\begin{figure}[h]
  \centering
  \includegraphics [width=1\columnwidth] {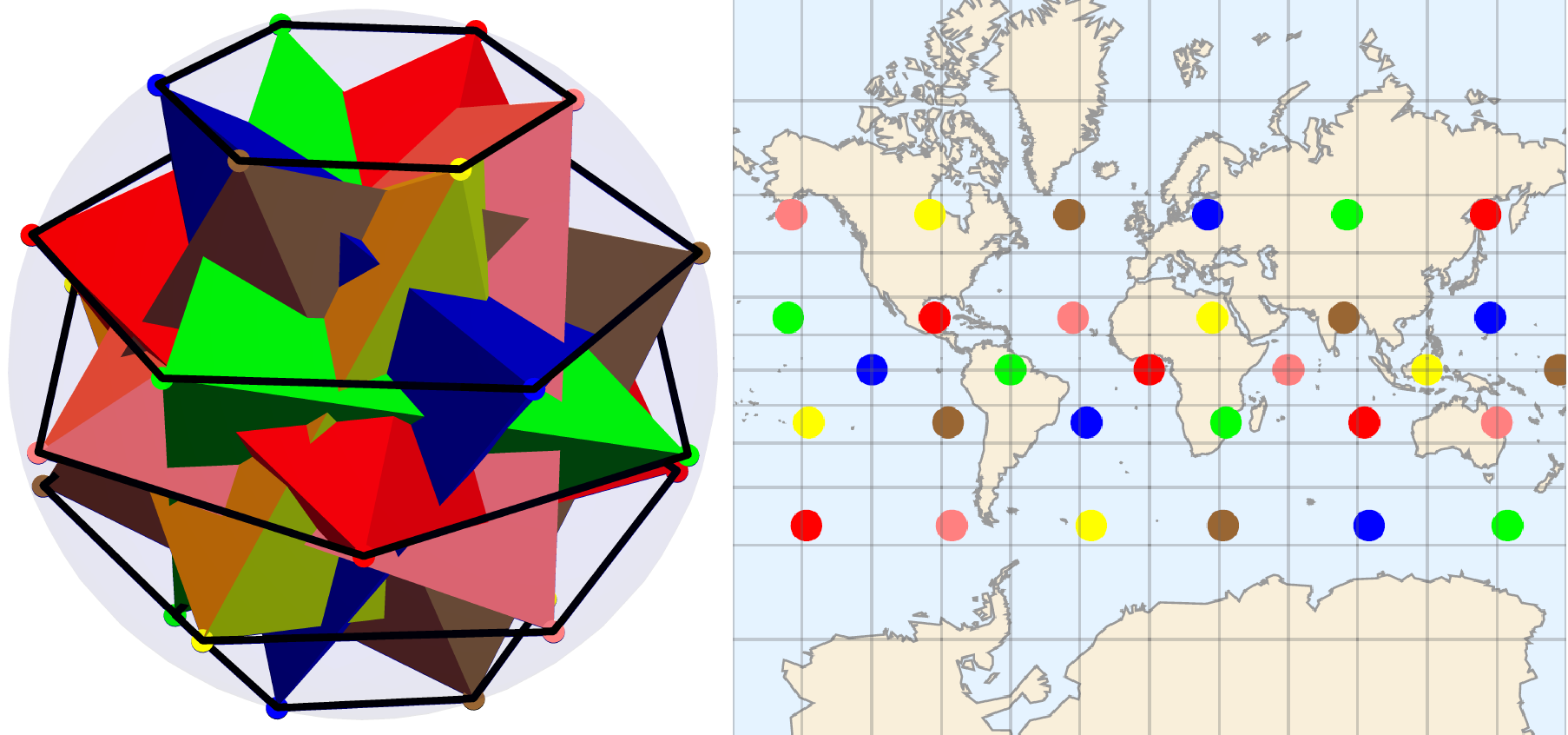}
  \caption{\small The selected most quantum basis in  $\mathcal{H}_6$ $(j=5/2)$, for which the sum of $\mathcal{B}_1$ and $\mathcal{B}_2$ achieves maximum. It is represented by $6\times5=30$ points. The sphere with a regular hexagon on each circle of latitude (left) and the Mercator projection (right).}
  \label{Dim-6-antykoh}
\end{figure}

The basis with an octahedral symmetry, found by the numerical search (see \cref{Dim-6-snopy}) is conjectured to be the most classical. All states in this basis are equivalent up to a rotation. By imposing this symmetry and orthogonality requirement one gets an analytical expression concerning the basis, for which the measure equals $\mathcal{B}_1(U^{c}_6)=8(137-34\sqrt{10})/2025\approx 0.116$. The basis contains the state
\begin{equation}
   \ket{\psi^{c}_6}=\mathcal{N}_3 \Bigl(\ket{\frac{5}{2},\frac{5}{2}}-\frac{1}{3}(2\sqrt{2}\text{-}\sqrt{5})\ket{\frac{5}{2},\text{-}\frac{3}{2}}\Bigr)
\end{equation}
and the remaining five states, which may be obtained by appropriate rotations, where $\mathcal{N}_3$ denotes the suitable normalization constant. Thus this basis belongs to the class of iso-coherent bases. In the Majorana representation, a state $\ket{\psi^{q}_6}$ is represented by a single star at the north pole and the remaining four equally distributed on a parallel with latitude defined by condition, $\tan\frac{\theta_7}{2}=(\frac{1}{3}(2\sqrt{10}-5))^{1/4}$, --~see red stars in \cref{Dim-6-snopy}. The corresponding unitary matrix reads
\begin{equation}
   \begin{aligned}
      U_6^{c}=\frac{1}{2}\left(\begin{matrix} 2a&0&\text{-}b&\text{-}b&\text{-}b&\text{-}b\\
0&2b&a&\text{-}a&\text{-}ia&ia\\
0&0&1&1&\text{-}1&\text{-}1\\
0&0&1&\text{-}1&i&\text{-}i\\
2b&0&a&a&a&a\\
0&2a&\text{-}b&b&ib&\text{-}ib\\\end{matrix}\right),
  \end{aligned}
  \label{U-6-max}
\end{equation}
where $a=\frac{1}{3}\sqrt{11/2+\sqrt{10}}$ and $b=(2-\sqrt{10})/6$.
\begin{conjecture}
The  most classical basis  in $\mathcal{H}_6$ is $U_6^{c}$ for which $\mathcal{B}_1=\frac{8(137-34\sqrt{10})}{2025}\approx 0.116$, --~see Eq. \cref{U-6-max}.
\end{conjecture}
Similarly to the case $N=5$, the numerical procedure leads to several solutions saturating the bound $\mathcal{B}_1\leq 1$. The standard method of increasing both $\mathcal{B}_1$ and $\mathcal{B}_2$ in a single step returns different bases, for which $\mathcal{B}_1$ and $\mathcal{B}_2<1$.
However, if we impose the maximization of the sum of measures $\mathcal{B}_1$ and $\mathcal{B}_2$, we observe the emergence of an intriguingly symmetric structure. This basis is comprised of 30 points that are equally distributed across five circles of latitude on a sphere, with each circle hosting six stars in vertices of a regular hexagon. The Majorana representation of this basis is displayed in \cref{Dim-6-antykoh}. A single state is represented, up to rotation, by a set of points in spherical coordinates $(\theta,\phi)$ given by
\begin{align}
    \nonumber\bigl\{(\pi/2,0), (\theta_8,5\pi/6-\delta_1), (\theta_9,-\pi/2-\delta_2), 
    \\(\pi-\theta_8,-5\pi/6+\delta_1), (\pi-\theta_9,\pi/2+\delta_2)\bigr\}.\label{Dim-6-min-sferyczne}
\end{align}
The remaining four states can be obtained through rotation around the $\hat{z}$ axis by an angle of $\pi/3$, so this basis is iso-coherent. Based on numerical experiments, we have imposed this parametrization and obtained numerical results for the four real parameters $\theta_8,\theta_9,\delta_1$, and $\delta_2$, resulting in one state described by Eq.~\cref{Dim-6-min-sferyczne} and four rotations. These rotations generate mutually orthogonal states up to accuracy of $10^{-8}$. Thus on basis, denoted as $\tilde{U}^{q}_6$, yields $\mathcal{B}_1\approx0.999994$ and $\mathcal{B}_2\approx0.989291$. It maximizes the sum of  $\mathcal{B}_1 + \mathcal{B}_2$, and leads to a solution with rotational symmetries. See \cref{Appendix B} for further details of the numerical procedure.

\subsection{Extremal basis for N=7}
\begin{figure}[h]
  \centering
  \includegraphics [width=1\columnwidth] {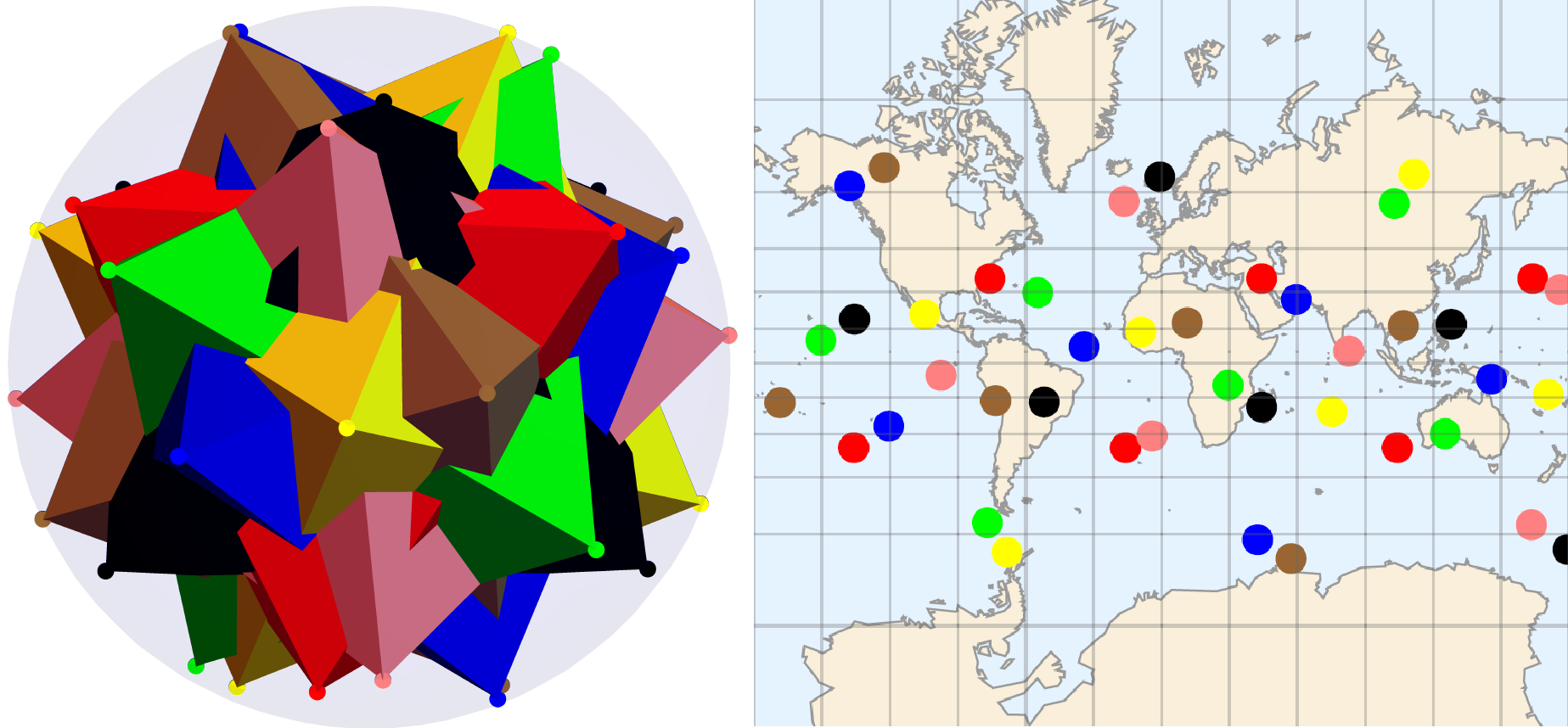}
  \caption{\small The most quantum basis in  $\mathcal{H}_7$ $(j=3)$ represented by $7\times6=42$ points. Each state is represented by a regular octahedron. The sphere (left) and the Mercator projection (right).}
  \label{Dim-7-osmiosciany}
\end{figure}
The most quantum basis $U_7^{q}$, found by numerical search is formed by seven regular octahedrons, --~see \cref{Dim-7-osmiosciany}. All states in this iso-coherent basis are equivalent up to a rotation of the state,
\begin{equation}
   \ket{\psi^{oct}_7}=\frac{1}{\sqrt{2}} (\ket{3,\text{-}2}+\ket{3,2}).
\end{equation}
No symmetry has been observed in the relative positions of those octahedrons. Further details regarding the numerics are provided in \cref{Appendix B}.

\begin{proposition}
The most quantum basis in $\mathcal{H}_7$ is represented by $U_7^{q}$ for which $\mathcal{B}_1=\mathcal{B}_2=\mathcal{B}_3=1$ and $\mathcal{B}_4=\frac{5}{6}$.
\end{proposition}

\subsection{Asymptotic results}
We observed that the average value of the measures of quantumness $\mathcal{B}_t(U)$ increase with the dimension $N$ of the Hilbert space $\mathcal{H}_N$ for any $t$ --~see \cref{wykres} as an example. 

In \cref{Appendix C} we prove the following statement.
\begin{theorem}\label{Theorem-srednia}
    The average value of $\mathcal{B}_t$ with respect to the Haar measure on $U(N)$ group is
    \begin{equation}
        \langle\mathcal{B}_t(U) \rangle_{U\in U(N)}=\frac{N-t-1}{N-t}.
    \end{equation}
\end{theorem}
Obtained results suggest that it is possible to construct a basis, using $t-$anticoherent states and $t$ grows with dimension $N$ of a Hilbert space. The overlap between spin coherent states as a function of their distance on a sphere goes to zero faster for higher spin states. It suggests that the minimal value of quantumness decreases with $N$. 
\begin{figure}[h]
  \centering
  \includegraphics [width=1\columnwidth] {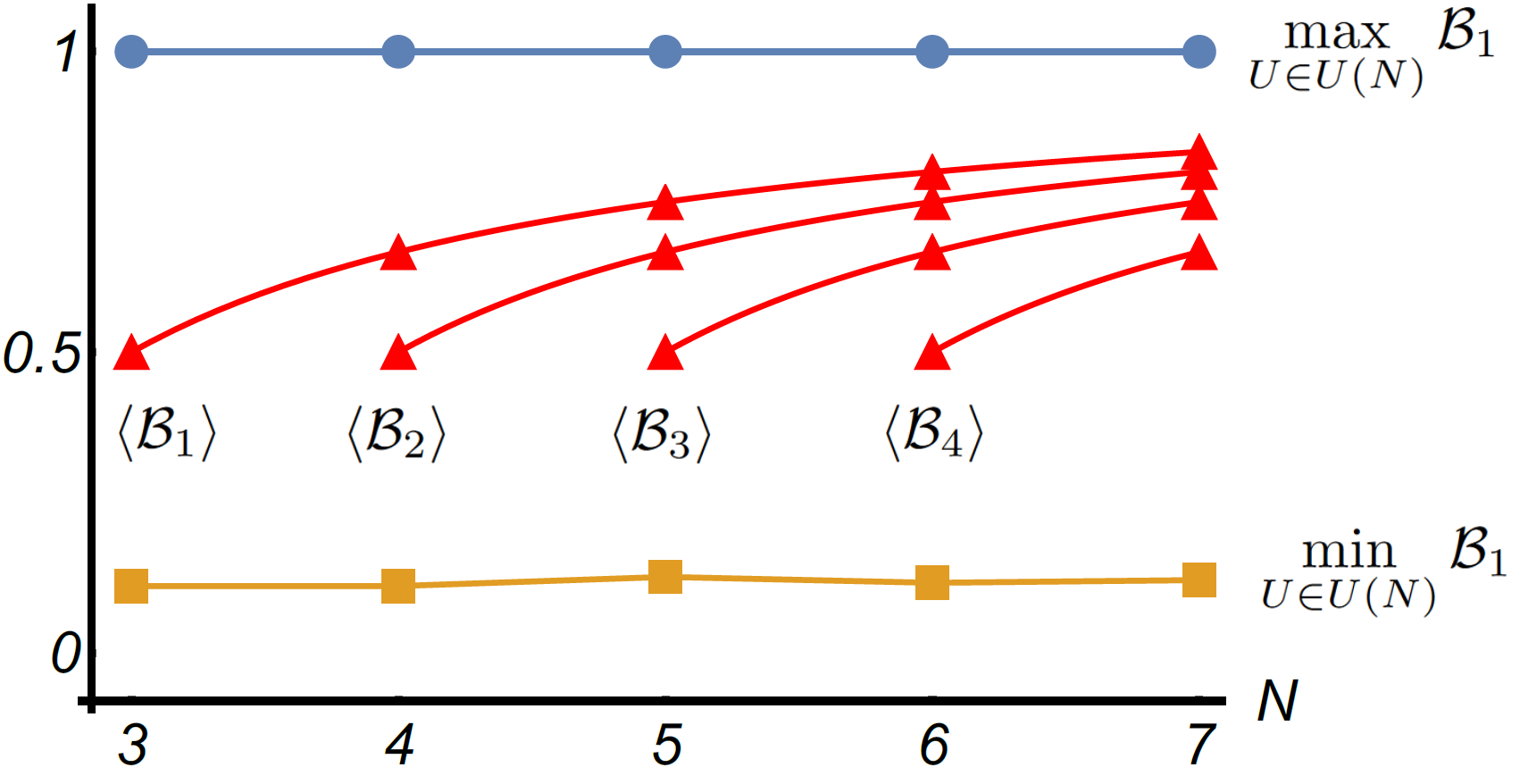}
  \caption{\small Maximal and minimal value of the quantumness $\mathcal{B}_1$, as a function of the dimension $N=2j+1$. Solid lines are plotted to guide the eye. The numerical average values $\langle\mathcal{B}_t\rangle$, obtained by averaging over Circular Unitary Ensemble (CUE), are denoted as the red triangles. Solid red lines show analytical result from Theorem 1.} 
  \label{wykres}
\end{figure}

\section{Husimi Function and Wehrl Entropy}\label{Wehrl}
Another useful measure of quantumness of a state $\ket{\psi}$ is its Wehrl entropy \cite{We78,Wehrl1979}, defined as an entropy of the quasi-probability distribution in a phase space. Suitable quasi-probability distribution is given by Husimi function (Q-function) \cite{Husimi}, defined as
\begin{equation}
    Q_{\psi}(\alpha)=|\braket{\psi|\alpha}|^2,
\end{equation}
where $\ket{\alpha}=\ket{\theta,\phi}$ is spin coherent state pointing in $(\theta,\phi)$ direction in spherical coordinates i.e. $\mathbb{C}\ni\alpha=\tan\frac{\theta}{2}e^{i\phi}$ and all stars representing this state in the stellar representation are placed in $(\theta,\phi)$ point on a sphere. Note, that points on a sphere (stars), representing a state $\ket{\psi}$ in the stellar representations are antipodal to zeros of the Husimi function $Q_{\psi}$ \cite{Zyczkowski}. The function $Q_{\psi}$ is normalized, bounded and always non-negative. 

For a pure spin state $\ket{\psi}\in\mathcal{H}_N$, the Wehrl entropy reads \cite{Wehrl1979},
\begin{equation}
    S_W(\ket{\psi})=-\frac{N}{4\pi}\int_{\Omega}d\Omega~Q_{\psi}(\alpha) \ln(Q_{\psi}(\alpha)),
\end{equation}
According to the long-standing Lieb conjecture \cite{Lee1988,Lieb_conjecture}, transformed into theorem by Lieb and Solovej \cite{Lieb_proof}, the Wehrl entropy is minimal for spin coherent states,
\begin{equation}
    S^{min}_W(\psi)=\frac{N-1}{N}=\frac{2j}{2j+1} \xrightarrow[N \to \infty]{} 1.
\end{equation}
Its maximal value is attained for states that are represented by stars equally distributed on a sphere (most quantum) \cite{Grassl}.

\onecolumngrid
\begin{center}
    \begin{table}[htb]
\centering
\setlength\extrarowheight{2pt}
\setlength{\tabcolsep}{2pt}
\begin{tabular}{|c|c|c|c|c|c|c|c|c|}
 \hline
&\text{Classical} &$\ket{j,m}$&\multicolumn{1}{c|}{Average }&\multicolumn{4}{c|}{Quantum} \\
 \hline
  N & $\mathcal{B}_1$ & $\mathcal{B}_1$  & $\langle\mathcal{B}_1(U) \rangle_{CUE}$ &$\mathcal{B}_1$ & $\mathcal{B}_2$& $\mathcal{B}_3$& $\mathcal{B}_4$  \\  
 \hline
 3 &  \color{purple}$\mathbf{\frac{1}{9}{\approx} 0.111}$  &\color{purple}$\mathbf{\frac{1}{3}{\approx} 0.333}$  &\color{purple} $\mathbf{\frac{1}{2}}$&\color{purple}$\mathbf{1}$ & - & -& - \\ 
 \hline
 4 & \color{blue}$\frac{1}{9}{\approx} 0.111$  &\color{purple}$\mathbf{\frac{4}{9}{\approx} 0.444}$ & \color{purple}$\mathbf{\frac{2}{3}{\approx}0.667}$ & \color{blue}1 & \color{blue}$3/4$& -& - \\
 \hline
 5 & $0.126$  &\color{purple} $\mathbf{1/2}$& \color{purple}$\mathbf{\frac{3}{4}=0.75}$& \color{blue}1 & \color{blue}1 & \color{blue}$2/3$& -\\
 \hline
 \multicolumn{1}{|c|}{\multirow{2}{*}{6}} & \multicolumn{1}{c|}{\multirow{2}{*}{\color{blue}$\frac{8(137\text{-}34\sqrt{10})}{2025}{\approx} 0.116$}}   &\multicolumn{1}{c|}{\multirow{2}{*}{\color{purple}$\mathbf{\frac{8}{15}{\approx} 0.533}$}}  &\multicolumn{1}{c|}{\multirow{2}{*}{  \color{purple}$\mathbf{\frac{4}{5}=0.8}$  }} &  $0.999994$ & $0.989$ & $0.879$ & $0.625$ \\
 \cline{5-8}
  & & &   &1 & 0.908 & 0.807& 0.625 \\
 \hline
  7 & - &\color{purple}$\mathbf{\frac{5}{9}{\approx} 0.556}$  & \color{purple}$\mathbf{\frac{5}{6}{\approx}0.833}$& 1 & 1 & 1 & 5/6 \\
 \hline
\end{tabular}
\caption{\label{Tabela}Measures of quantumness $\mathcal{B}_t$ for identified extremal bases of order $N=2j+1=3,\ldots,7$, canonical spin $j$ bases $\ket{j,m}$ \cite{Kz01} and average over unitary group with respect to Circular Unitary Ensemble (CUE) $\langle\mathcal{B}_1(U) \rangle_{CUE}$. Analytical results are shown in bold purple, while analytical results with constraints suggested by numerical outcomes are shown in blue. The remaining bases were obtained via numerical search with imposed symmetry constraints to enhance precision. Analyzing the most quantum basis in $N=6$, we identified two results, the first with the symmetric structure (see \cref{Dim-6-antykoh}) and the second with the highest measure of quantumness $\mathcal{B}_1$ without any specific symmetries.}
\end{table}
\end{center}
\twocolumngrid

We investigated the problem of finding bases of extreme mean Wehrl entropy of states forming the basis. We use a similar algorithm as for the measure $\mathcal{B}_t$ and obtain the same extremal bases (except for the case of the most quantum basis in $N=6$) as for the previous measure. In search for the most quantum basis in $N=6$ the algorithm did not converge to a single local extremum, similarly to the case of maximizing $\mathcal{B}_1$ discussed in \cref{Main_results}.

An average value of the Wehrl entropy of random pure state equals \cite{SZ98}
\begin{equation}
  \langle S_W(\ket{\psi}) \rangle_{\ket{\psi}\in \mathcal{H}_N}= \sum^{N}_{k=2}\frac{1}{k},  
\end{equation}
and gives the mean value for a generic basis corresponding to a Haar random unitary matrix $U$. 
All results for the Wehrl entropy are presented in the \cref{Tabela2}.

\begin{figure}[h!]
  \centering
  \includegraphics [width=1\columnwidth] {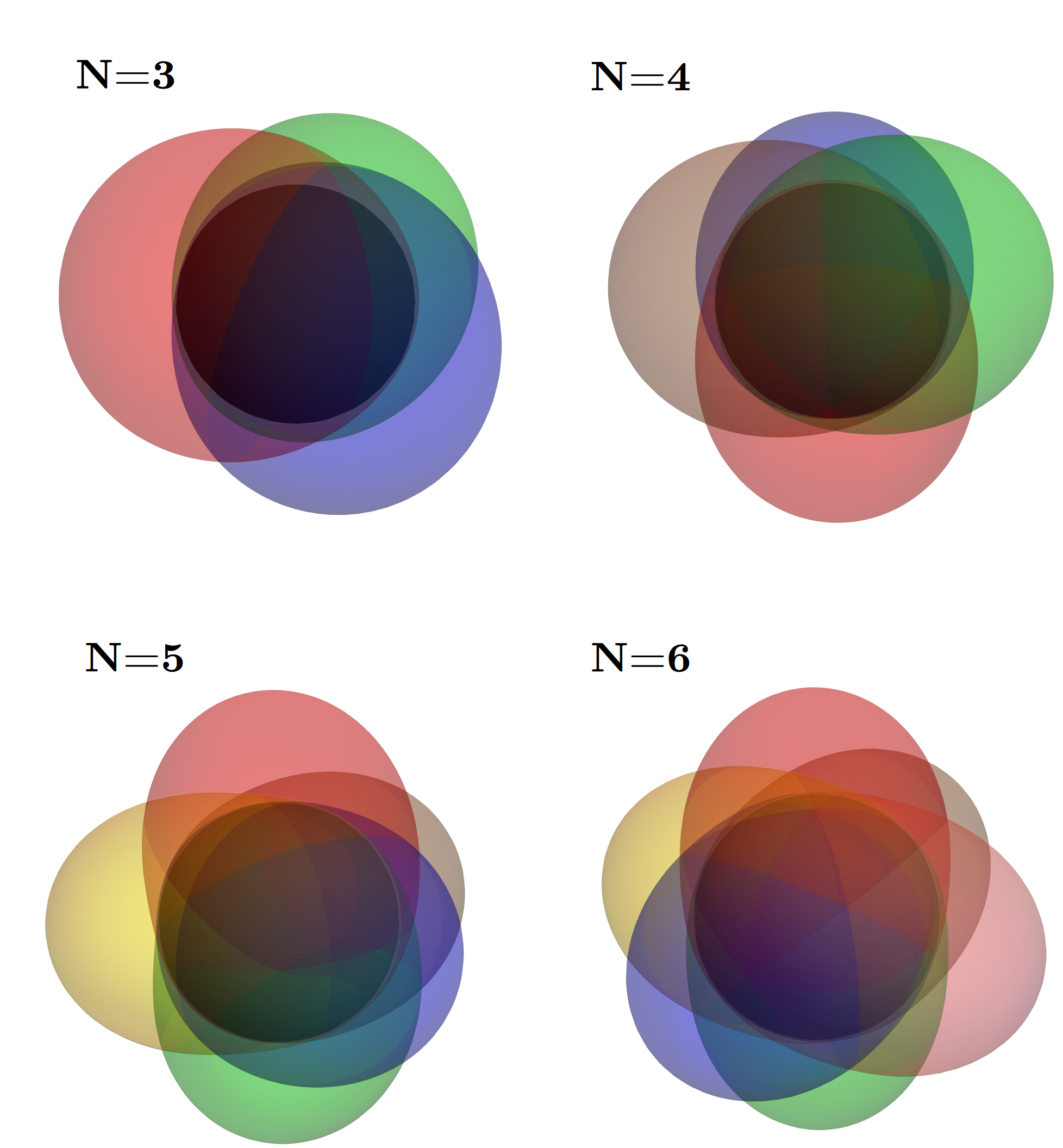}
  \caption{\small The Husimi functions $Q_{\psi_i}$ of vectors $\ket{\psi_i}$, $i=1,\ldots,N$ of the most classical bases in $N=3,4,5,6$ as envelopes on the sphere. Black shadow denotes the position of the Bloch sphere. Colors of the Husimi function correspond to the colors of stars in \cref{j=1-coh-basis,Dim-4-snopy,Dim-5-snopy,Dim-6-snopy}. The maximum of the red distribution is placed antipodal to the cluster of Majorana stars, where the red point is missing.  }
  \label{AllBasesHusimi}
\end{figure}

Plots of Husimi functions $Q_{\psi_i}$ of states forming the most classical bases are presented in \cref{AllBasesHusimi}. Observe the relation to the constellations of Majorana stars shown in \cref{j=1-coh-basis,Dim-4-snopy,Dim-5-snopy,Dim-6-snopy}. For any $N$ the stars form $N$ clusters containing $N-1$ points each. The corresponding Husimi function $Q_{\psi_i}$ is localized at the point antipodal to the cluster in which a given color is missing. Shapes of states with high quantumness are delocalized and are hardly distinguishable, so will not be reproduced here.

\onecolumngrid
\begin{center}
\begin{table}[htb]
\centering
\setlength\extrarowheight{2pt}
\setlength{\tabcolsep}{1pt}
\begin{tabular}{|c|c|c|c|c|}
 \hline
&\text{Classical} &$\ket{j,m}$&\multicolumn{1}{c|}{Average}&\text{Quantum} \\
 \hline
  N  & $\bar{S}_W$ &   $\bar{S}_W$ &   $\langle \bar{S}_W(U)\rangle_{CUE}$&  $ \bar{S}_W$ \\  
 \hline
 3 &  0.712 & \color{purple}$\mathbf{1\text{-}\frac{\ln2}{3}{\approx} 0.769}$ &\color{purple}$\mathbf{\frac{5}{6}{\approx}0.833}$& \color{blue}$\frac{5}{3}\text{-}\ln2{\approx}0.974$\\ 
 \hline
 4  &0.831  &\color{purple}$\mathbf{\frac{3}{2}\text{-}\frac{\ln3}{2}{\approx} 0.951}$  &\color{purple}$\mathbf{\frac{13}{12}{\approx}1.08}$  &1.24\\
 \hline
 5 & 0.912 &\color{purple}$\mathbf{2\text{-}\frac{\ln96}{5}{\approx} 1.09}$& \color{purple}$\mathbf{\frac{77}{60}{\approx}1.28}$  &1.50\\
 \hline
  6  & 0.965 &\color{purple}$\mathbf{\frac{5}{2}\text{-}\frac{1}{3}\ln50{\approx} 1.20}$ &\color{purple}$\mathbf{\frac{29}{20}=1.45}$  &1.65\\
 \hline
  7  & - &\color{purple}$\mathbf{3\text{-}\frac{\ln162000}{7}{\approx}1.29}$  &\color{purple}$\mathbf{\frac{223}{140}{\approx}1.59}$  &1.84\\
 \hline
\end{tabular}
\caption{\label{Tabela2}
Means Wehrl entropy $\bar{S}_W $ of vectors forming the extremal bases of order $N=3,\ldots,7$, on canonical $J_z$ basis $\ket{j,m}$ with $j=(N-1)/2$ and average over unitary group \cite{SZ98}. Analytical results are shown in bold purple, while analytical results with constraints suggested by numerical outcomes are shown in blue. The remaining bases were obtained via numerical search with imposed symmetry constraints to enhance precision.}
\end{table}
\end{center}
\twocolumngrid

Another interesting measure of quantumness is the maximum of the Husimi function for a state $\ket{\psi}\in\mathcal{H}_N$,
\begin{equation}
\label{eq:maxHusismi}
    Q_{max}(\psi)=\max_{\alpha\in\mathbb{C}}|\braket{\alpha|\psi}|^2.
\end{equation}
The $Q_{max}$ value determines the minimal Fubini-Study distance of the state analyzed to the manifold ~of ~spin-coherent states, ~which ~reads $D_{FS}=\arccos{\sqrt{Q_{max}}}$ \cite{Zyczkowski}. Therefore, states $\ket{\psi}$ that minimize $Q_{max}(\psi)$ are considered as most quantum \cite{GiraudQueens,Grassl}.  
Interestingly, finding extremes of the maximum of the Husimi function in small dimensions leads to another set of bases than those that are extreme for anticoherence and Wehrl entropy. Indeed, we used a similar search algorithm as for the measure $\mathcal{B}_t$ to find a basis with the extreme values of (\ref{eq:maxHusismi}). However, the $Q_{max} $ measure seems to be problematic for the numerical search, as in many cases it has a numerical plateau around its extreme values. In such cases, the numerical search for a basis with the extreme values of $Q_{max}$ leads to bases without any particular symmetries. In some special cases (bases in dimension $N=3$, and the most coherent basis in dimensions $N=5,6$), the $Q_{max} $ extreme basis coincides with extreme $\mathcal{B}_t$ (anticoherent) basis. Due to the aforementioned problem of finding the extreme value of $Q_{max}$, and the lack of symmetry in a numerical approximation of such a solution, we do not present extreme bases for the function $Q_{max}$ in a detailed way. \cref{Tabela3} presents the values of  $Q_{max}$ measure for the bases described in \cref{Main_results}. Notice that due to the previously mentioned problem, in many cases these bases do not reach the extreme values of the $Q_{max}$ function.

\section{Permutation symmetric states of multiqubit systems}\label{Permutation-sym-states}

Due to the correspondence \cref{Identification} between spin-$j$ states and symmetric states of $2j$ spin-$1/2$ particles and the fact that the measure $\mathcal{A}_1$ of 1-anticoherence defined in  Eq.~\cref{anticoherence} is equivalent to the linear entropy of reduced density operators, the extremal bases we found correspond to the most and the least entangled bases with respect to the bipartition into one and $2j-1$ qubits. Specifically, the 1-anticoherent states, $\mathcal{A}_1(\ket{\psi})=1$, are maximally entangled, while the coherent states $\mathcal{A}_1(\ket{\psi})=0$ represent biseparable states of the composite system. 

Note that the basis \cref{U-3-min}, represented in \cref{j=1-ant-basis} by a regular octahedron, is equivalent to the well-known Bell basis in the symmetric sector of $\mathcal{H}^{\otimes 2}_2$,
\begin{align}
\nonumber
       \ket{\psi_1}&=\frac{1}{\sqrt{2}}(\ket{01}+\ket{10}),
\\
\label{Bell}
       \ket{\psi_2}&=\frac{1}{\sqrt{2}}(\ket{00}+\ket{11}),
\\
\nonumber       
       \ket{\psi_3}&=\frac{1}{\sqrt{2}}(\ket{00}-\ket{11}).
    \end{align}
The last state that completes the orthonormal basis in $\mathcal{H}^{\otimes 2}_2$ is the singlet state 
\begin{equation}
    \ket{\psi_4}=\frac{1}{\sqrt{2}}(\ket{01}-\ket{10}),
\end{equation}
which does not belong to the symmetric subspace.
 In a similar way, the most quantum basis in $\mathcal{H}_4$, represented by a cuboctahedron shown in \cref{Dim-4-anticoherent}, corresponds to the maximally entangled basis in the symmetric subspace of $\mathcal{H}^{\otimes 3}_2$. Each state is equivalent, up to a local unitary transformation $U(2)\otimes U(2)\otimes U(2)$, to a $GHZ$ state \cite{Zyczkowski}, 
 \begin{equation}
     \ket{GHZ}=\frac{1}{\sqrt{2}}(\ket{000}+\ket{111}).
 \end{equation}
Therefore, this basis is iso-entangled \cite{Iran,Grzegorz,Kuba, SCZG23}.

 In $\mathcal{H}_5$, the most quantum basis \cref{U-5-min}, represented in \cref{Dim-5-dodecahedron} by five regular tetrahedrons that form a regular dodecahedron, corresponds to the iso-entangled basis in the symmetric sector of $\mathcal{H}^{\otimes 4}_2$. This basis consists five, maximally entangled states of four qubits, which are known to be the most sensitive states for alignment of the reference frame \cite{Hyllus} and having the highest geometric entanglement \cite{MartinMultiqubit}. A simple calculation leads to the form
 \begin{align}
     \ket{\psi_{tet.}}=\frac{1}{\sqrt{3}}\ket{0000}+\frac{1}{\sqrt{6}}(\ket{0111}
     \label{Tetrahedron-symmetric}
     \\\nonumber+\ket{1011}+\ket{1101}+\ket{1110}).
 \end{align}
 In a similar way the matrix $U_7^{q}$ of maximal quantumness leads to the iso-entangled basis of the symmetric sector of $\mathcal{H}^{\otimes 6}_2$ with the maximal geometric entanglement. Each state is equivalent, up to a local unitary, to 
  \begin{equation}
       \ket{\psi_{oct.}}=\frac{1}{\sqrt{2}}(\ket{D_{6,1}}-\ket{D_{6,5}}),
     \label{Octahedron-symmetric}
  \end{equation}
which is known to display the highest geometric entanglement \cite{MartinMultiqubit}.

\section{Concluding remarks}\label{Concluding-Remarks}


This study introduces the measure $\mathcal{B}_t$ as the tool to characterize the quantumness of a given basis in $\mathcal{H}_N$. The search for the most quantum bases for $N=3,4,5$ and $7$ is performed. Numerical results suggest, that the obtained solutions are unique.  A set of candidates for the "classical" bases consisting of the most spin-coherent states is indicated for $N=3,4,5,6$. Presented bases give also extremal values of the average Wehrl entropy of basis vectors. Some of the most quantum bases, analyzed in the stellar representation of Majorana, reveal symmetries of Platonic solids. Most classical bases display intriguing structures. Stars of different colors group in $N$ clusters consisting of exactly $N-1$ stars. In each cluster, only a single color is missing and the Husimi function corresponding to this color is concentrated at the point antipodal to the barycenter of the cluster - see \cref{AllBasesHusimi}. Obtained results, including the average and the extremal values of $\mathcal{B}_1$, are presented in \cref{wykres} and summarized in the \cref{Tabela}. 

Extremal states, coherent and anticoherent, have practical applications in quantum metrology as optimal rotosensors \cite{Kolenderski,Chryssomalokos,GoldbergOptimal,MartinOptimal}. This work provides a natural extension of previous studies concerning the search for such states proposing optimal orthogonal measurements of Lüders and von Neumann of the extreme spin coherence. Note, that several of the presented bases have specific rotational symmetries and are iso-coherent which allows one to obtain the entire basis by rotation of a reference state and therefore makes them easier to prepare. Such states, generating basis by rotation, are called optimal quantum protractors \cite{Kamil}. Observe that all bases found, except the basis for $N=5$, are generated by a single quantum protractor. Therefore t-anticoherence measure $\mathcal{A}_t$ of all states in a basis are equal to the quantumness measure $\mathcal{B}_t$ of the entire basis, which are iso-coherent.

\onecolumngrid
\begin{center}
\begin{table}[htb]
\centering
\setlength\extrarowheight{2pt}
\setlength{\tabcolsep}{1pt}
\begin{tabular}{|c|c|c|c|c|}
 \hline
&\text{Classical} &$\ket{j,m}$&\multicolumn{1}{c|}{Average}&\text{Quantum} \\
 \hline
  N  & $\bar{Q}_{max}$&   $\bar{Q}_{max}$ &   $\langle \bar{Q}_{max}\rangle_{CUE}$&  $\bar{Q}_{max}$ \\  
 \hline
 3 &  \color{purple}$\mathbf{\frac{1}{6} \left(3+2 \sqrt{2}\right)\approx 0.971}$ & \color{purple}$\mathbf{\frac{5}{6}{\approx 0.833}}$ &0.831  & \color{purple}$\mathbf{\frac{1}{2}}$\\ 
 \hline
 4  &\color{purple}$\mathbf{\frac{1}{6} \left(3+2 \sqrt{2}\right)\approx 0.971}$  &\color{purple}$\mathbf{\frac{13}{18}{\approx}0.722}$  & 0.718  & \color{purple}$\mathbf{\frac{1}{2}}$\\
 \hline
 5 & $\frac{2*0.9518+3*0.9608}{5}\approx 0.957$ & \color{purple}$\mathbf{\frac{556403}{1179648}{\approx}0.472}$&  0.633  & \color{purple}$\mathbf{\frac{1}{3}}$\\
 \hline
 \multicolumn{1}{|c|}{\multirow{2}{*}{6}} & \multicolumn{1}{c|}{\multirow{2}{*}{\color{purple}$\mathbf{\frac{1}{18} \left(11+2 \sqrt{10}\right)\approx 0.962}$}} &\multicolumn{1}{c|}{\multirow{2}{*}{\color{purple}$\mathbf{\frac{1097}{1875}{\approx}0.585}$}} & \multicolumn{1}{c|}{\multirow{2}{*}{    0.568     }} &0.331\\
 \cline{5-5}
  & & & & 0.333\\
 \hline
  7  & - &\color{purple}$\mathbf{\frac{1223}{2268}{\approx}0.539}$  & 0.514 & \color{purple}$\mathbf{\frac{2}{9}{\approx}0.222}$\\
 \hline
\end{tabular}
\caption{\label{Tabela3}
Arithmetic means of the $Q_{max}$ function for states in identified extremal bases with respect to $\mathcal{B}_t$ measure for dimensions $N=3,\ldots,7$, canonical spin $j$ bases $\ket{j,m}$ and numerical average over Circular Unitary Ensemble (CUE). Analytical results are shown in bold purple.}
\end{table}
\end{center}
\twocolumngrid

We also considered other measures of quantumness of vectors forming a given basis. Optimization of the mean Wehrl entropy of $N$ orthogonal vectors leads to the same bases distinguished by extremal values of the quantities $\mathcal{B}_t$, with a single exception of the quantum basis for $N=6$ --~see \cref{Tabela2}. Similar bases are distinguished by the search for the extremal values of the maximum $\bar{Q}_{max}$ of the Husimi function, averaged over all orthogonal states of the basis --~see \cref{Tabela3}. This quantity characterizes the average distance of the basis vectors $\ket{\psi_i}$ to the set of spin coherent states. In such a way, we presented in this work several alternative possibilities to distinguish orthogonal measurements of $N$-dimensional state with extremal properties concerning the minimal, maximal and average spin coherence of the basis vectors.
\section{Acknowledgements}\label{Acknowledgements}
We would like to express our gratitude to Rafa{\l} Bistro{\'n} and Jakub Czartowski for engaging in discussions that have enriched this research significantly. We are also thankful to Grzegorz Rajchel-Mieldzio{\'c} for constructive remarks. Financial support provided by the Foundation for Polish Science through the TEAM-NET project number POIR.04.04.00-00-17C1/18-00 and by the Narodowe Centrum Nauki provided through the Quantera project number 2021/03/Y/ST2/00193 is gratefully acknowledged. A.~B.~acknowledges support by an NWO Vidi grant Project No. VI.Vidi.192.109.

\onecolumngrid

\appendix
\section{Parametrization of bases in $\mathcal{H}_3$}\label{Appendix A}
 
Any orthonormal basis in $\mathcal{H}_3$, consisting of three states $\ket{\Psi_3}$, $\ket{\Psi^{\prime}_3}$ and $\ket{\Psi^{\prime\prime}_3}$ can be expressed, up to a $SU(2)$ rotation, in terms of three real parameters $\Theta_1, \Theta_2\in [0,\pi)$ and $\Phi\in[0,2\pi)$,
\begin{align}
\nonumber
       \ket{\Psi_3}&=\mathcal{N}_4(\ket{1,1}-\tan^2\frac{\Theta_1}{2}\ket{1,\text{-}1}),~~~~~~
\\
\label{parametryzacja-H3}
       \ket{\Psi^{\prime}_3}&=\mathcal{N}_5(\ket{1,1}+\Upsilon\ket{1,0}+\cot^2\frac{\Theta_1}{2}\ket{1,\text{-}1}),
\\
\nonumber       
       \ket{\Psi^{\prime\prime}_3}&=\mathcal{N}_6(\ket{1,1}-\frac{1+\cot^4\frac{\Theta_1}{2}}{\Upsilon^{\ast}}\ket{1,0}+\cot^2\frac{\Theta_1}{2}\ket{1,\text{-}1}),
    \end{align}
where $\Upsilon=(e^{-i\Phi}\cot\frac{\Theta_2}{2}\cot^2\frac{\Theta_1}{2}+e^{i\Phi}\tan\frac{\Theta_2}{2})/\sqrt{2}$, and $\mathcal{N}_4$, $\mathcal{N}_5$, $\mathcal{N}_6$ denote suitable normalization constants.  In the stellar representation of Majorana, the state $\ket{\Psi_3}$ is represented by two points on the opposite sides of a single circle of latitude, $\Theta_1$. The state $\ket{\Psi_3}$ was chosen arbitrarily without loosing generality, and the states $\ket{\Psi^{\prime\prime}_3}$ and $\ket{\Psi^{\prime}_3}$ were obtained by imposing orthogonality conditions.

\section{Numerical results}\label{Appendix B}
 
In this Section, we present further details concerning the numerical procedure. Detailed expressions for the vectors forming orthonormal bases found numerically are available online \cite{strona}. 
\subsection*{$\mathbf{N=5}$}
The least quantum basis in $\mathcal{H}_5$, for which $\mathcal{B}_1\approx0.1263012$, is generated by rotation of two reference states \cref{Dim-5-max-stan1} and \cref{Dim-5-max-stan2}
with parameters $\chi=(r_2+1/r_2+2\cos\phi)/2$ and $r_1=\sqrt[3]{4/(1/r^{2}_2+r_{2}^2+2\cos\phi)}$.
Numerical optimization gives the following ~values ~of ~the ~free ~parameters, $~r_2\approx7.564405$ ~and $~\phi\approx0.93380835$.
The basis is orthonormal up to accuracy $\sum_{i,j,k}|U^{\ast}_{ij}U_{ik}-\delta_{jk}|\approx2.1\times10^{-15}$.
\subsection*{$\mathbf{N=6}$}
Minimizing $\mathcal{B}_1$ leads to several different bases with $\mathcal{B}_1\approx1$, without any internal symmetry. To distinguish among them, we analyzed the sum of $\mathcal{B}_1 + \mathcal{B}_2$. The measure $\mathcal{B}_1$ is no longer $\approx1$, but this choice leads to a rotational symmetry by $\pi/3$ and an internal state symmetry \cref{Dim-6-min-sferyczne}. Numerical search gave us the following values of the parameters of the state \cref{Dim-6-min-sferyczne}, $\theta_8\approx0.5922575$, $\theta_9\approx1.1820735$, $\delta_1\approx0.0822441$, $\delta_2\approx0.0522207$.
Therefore, the corresponding reference vector in $\mathcal{H}_6$ reads,
\begin{align*}
    \ket{\psi^{q}_6}\approx(0.4082482, -0.3757166 - 
 0.1596989 i, 0.0850774 - 
 0.3992850 i, \\0.0850774 - 
 0.3992850 i, -0.3757166 - 
 0.1596989 i, 0.4082483).
\end{align*}
The other four vectors are obtained by rotation of $\ket{\psi^{q}_6}$ around the $\hat{z}$ axis by an angle of $\pi/3$. Rotation matrix reads $\hat{R}_{\hat{z}}(\pi/3)=\diag(1,e^{i\pi/3},e^{i2\pi/3},e^{i\pi},e^{i4\pi/3},e^{i5\pi/3})$ and the entire iso-coherent basis takes the form, 
\begin{align*}
    \bigl\{\hat{R}_{\hat{z}}(\pi/3)^k\ket{\psi^{q}_6} \bigr\}^{5}_{k=0}.
\end{align*}
The obtained constellation of vectors provides the following values of the quantumness measures, $\mathcal{B}_1\approx0.9999940$, $\mathcal{B}_2\approx0.9892914$ and $\mathcal{B}_3\approx0.8793702$. The basis is orthonormal up to accuracy $\sum_{i,j,k}|U^{\ast}_{ij}U_{ik}-\delta_{jk}|\approx9.4\times10^{-7}$.
\subsection*{$\mathbf{N=7}$}
The basis $U^{q}_7$  is orthonormal with accuracy $\sum_{i,j,k}|U^{\ast}_{ij}U_{ik}-\delta_{jk}|\approx6.3\times10^{-15}$. Positions of stars representing states in this basis differ from those of the regular octahedrons by less than $10^{-15}$, for a unit radius of the sphere.

\section{Anticoherence measure in terms of matrix elements.}\label{Appendix C} 
The t-anticoherence measure $\mathcal{A}_t(\ket{\psi_i})$ of a state $\ket{\psi_i}$ is defined in Eq.~\cref{anticoherence} using purity of the reduced state $F(\ket{\psi_i})$, discussed in \cite{GiraudTensor,Baguette-measure}. Thus, the quantumness measure $\mathcal{B}_t(U)$ of an orthonormal basis $U=\{\ket{\psi_1},\ldots,\ket{\psi_N}\}$, can be expressed in terms of the purity of the corresponding reduced symmetric states $F(\ket{\psi_i})$. The purity $R_t$ of the state $F(\ket{\psi_i})$, corresponding to  $\ket{\psi_i}$, expressed in the eigenbasis \cref{state-ang-mom-bas} of angular momentum operator $J_z$ reads,
\begin{equation}
\label{R}
  R_t(\ket{\psi_i})=\sum_{k_1=0}^{t}\sum_{k_2=0}^{t}\bigg|\sum_{k=-j}^{j\text{-}t}Z^{*}_{j\text{-}k\text{-}k_1}Z_{j\text{-}k\text{-}k_2}\Gamma_{j+k}^{k_1k_2}\bigg|^2,
\end{equation}
where 
\begin{equation}
\label{Gamma}
    \Gamma_{k}^{k_1k_2}=\frac{1}{C^{2j}_{t}}\sqrt{C^{k+k_1}_{k}C^{2j\text{-}k\text{-}k_1}_{t\text{-}k_1}C^{k+k_2}_{k}C^{2j\text{-}k\text{-}k_2}_{t\text{-}k_2}}
\end{equation}
and $C^l_q=\binom{l}{q}$ if $0\le q\le l$ and 0 otherwise. Then the t-coherence measure $\mathcal{B}_t(U)$ of an orthonormal basis, represented by unitary matrix $U$ can be expressed in terms of its matrix elements $U_{ki}=\braket{j,j+1-k|\psi_i}$ as
\begin{equation}
    \mathcal{B}_t(U)=\frac{t+1}{N t}\sum_{p=1}^{N}\biggl(1-\sum_{k_1,k_2=0}^t\biggl|\sum_{k=-j}^{j-t}U^{\ast}_{j-k-k_1+1,p}U_{j-k-k_2+1,p}\Gamma^{k_1k_2}_{j+k}\biggr|^2\biggr).\label{anticoherence-U}
\end{equation}

\subsection*{Proof of the Theorem 1}
Expanding Eq.~\cref{anticoherence-U} gives the sum of expressions of the structure
\begin{equation}
    U^{\ast}_{j-k-k_1+1,p}U_{j-k-k_2+1,p}U_{j-k'-k_1+1,p}U^{\ast}_{j-k'-k_2+1,p}\Gamma^{k_1k_2}_{j+k}\Gamma^{k_1k_2}_{j+k'},
\end{equation}
that could be averaged over unitary group $U(N)$ by the Weingarten calculus \cite{Weingarten}
\begin{equation}
    \begin{aligned}
        \int_{U(N)} dU U_{ij}U_{kl}U^{\ast}_{mn}U^{\ast}_{pq}=(\delta_{im}\delta_{jn}\delta_{kp}\delta_{lq}+\delta_{ip}\delta_{jq}\delta_{km}\delta_{ln})\text{Wg}(1^2,N)\\ +(\delta_{im}\delta_{jq}\delta_{kp}\delta_{ln}+\delta_{ip}\delta_{jn}\delta_{km}\delta_{lq})\text{Wg}(2,N),
    \end{aligned} 
\end{equation}
where $\text{Wg}$ is the Weingarten function. Inserting the known values of the Weingarten functions (see \cite{Collins}) into Eq.~\cref{anticoherence-U}, using expression \cref{Gamma} and applying the following identity
\begin{equation}
    \sum^{p}_{m=0}\binom{m}{l}\binom{p-m}{k-l}=\binom{p+1}{k+1},
\end{equation}
one arrives at the desired expression
\begin{equation}
        \langle\mathcal{B}_t(U) \rangle_{U(N)}=\frac{N-t-1}{N-t}.
    \end{equation}
\hspace{15cm}\qedsymbol{}
\section{Extremality of solution.}\label{Appendix D}

Extremal point $\mathbf{x_0}$ of n-variable function $f$ satisfies $\mathbf{\nabla}f|_{\mathbf{x_0}}=\mathbf{0}$. The most convenient way to determine whether it is a local minimum, maximum, or saddle point is the analysis of the positive definiteness of the Hessian $H_f$ defined as,
\begin{equation}
   \begin{aligned}
     H_f=\left(\begin{matrix} \frac{\partial^2f}{\partial x_1^2}&\frac{\partial^2f}{\partial x_1\partial x_2}&\ldots&\frac{\partial^2f}{\partial x_1\partial x_n} \\
\vdots&\vdots&\ddots&\vdots\\
\frac{\partial^2f}{\partial x_n\partial x_1}&\frac{\partial^2f}{\partial x_n\partial x_2}&\ldots&\frac{\partial^2f}{\partial x_n^2} \\\end{matrix}\right).
  \end{aligned}
  \label{Hessian}
\end{equation}
\begin{enumerate}
\item If $H_f|_{\mathbf{x}=\mathbf{x_0}} > 0\iff \mathbf{x_0}$ is a local maximum.
\item If $H_f|_{\mathbf{x}=\mathbf{x_0}} < 0\iff \mathbf{x_0}$ is a local minimum.
\item If $H_f|_{\mathbf{x}=\mathbf{x_0}} = 0\iff \mathbf{x_0}$ is a saddle point.
\end{enumerate}

Following \cite{Rajchel} we recall the Lie group structure of the manifold of unitary matrices to introduce directions in the neighborhood of the matrix U. Lie algebra of unitary matrices is formed by Hermitian matrices. We select the following basis,
\begin{equation}
    \begin{aligned}
    &H_{ii}\text{=}\ket{i}\bra{i} ~~\text{for} ~~i\in{1,\ldots,N},
    \\&H^+_{kl}\text{=}\ket{k}\bra{l}+\ket{l}\bra{k} ~~\text{for}~ k,l\in{1,\ldots,N}, ~k\neq l,
    \\&H^-_{kl}\text{=}i(\ket{k}\bra{l}- \ket{l}\bra{k}) ~~\text{for}~ k,l\in{1,\ldots,N}, ~k\neq l,
    \end{aligned}\label{Lie-algebra}
\end{equation}
that give $N^2$ directions in the neighbourhood of any $U$. Now we may define the derivative of a function $f$ on a unitary matrix as
\begin{equation}
    \nabla_rf(U)=\lim_{\epsilon_r\rightarrow 0}\frac{f(U_{\epsilon_r})-f(U)}{\epsilon_r},
\end{equation}
where $U_{\epsilon_r}$ is the matrix $U$, transformed in $r$-th direction as $U_{\epsilon_r}=U\exp(i\epsilon_r H^r)=U(\mathbb{1} +i\epsilon_r H_r+ O(\epsilon^2_r))$, where $H^r$ is an element of the Lie algebra \cref{Lie-algebra}.

For a given unitary matrix $U$ the derivative of the quantumness $\mathcal{B}_t(U)$, determined by Eqs.~\cref{anticoherence-measure,anticoherence-U} reads,
\begin{equation}
     \begin{aligned}
     \nabla_r\mathcal{B}_t(U)=
       -4\frac{t+1}{N t}\sum_{p,p'=1}^{N}\sum_{k_1,k_2=0}^{t}\sum_{k,k'=-j}^{j-t}\Gamma_{j+k}^{k_1k_2}\Gamma_{j+k'}^{k_1k_2}\text{Im}\Bigl(U_{j-k'-k_1+1,p}U^{\ast}_{j-k'-k_2+1,p}\times
       &\\\bigl(U^{\ast}_{j-k-k_1+1,p}U_{j-k-k_2+1,p'} H^{r\ast}_{p',p} -U_{j-k-k_2+1,p}U^{\ast}_{j-k-k_1+1,p'} H^{r\ast}_{p',p}   \bigr)\Bigr),\hspace{0.3cm}
    \end{aligned}
\end{equation}
where $\Gamma_{k}^{k_1k_2}$ is defined in Eq.~\cref{Gamma}. In a similar way, we arrive at the second derivative:
\begin{equation}
     \begin{aligned}
     \nabla_r\nabla_s\mathcal{B}_t(U)=
       -8\frac{t+1}{N t}\sum_{p,p',p''=1}^{N}\sum_{k_1,k_2=0}^{t}\sum_{k,k'=-j}^{j-t}\Gamma_{j+k}^{k_1k_2}\Gamma_{j+k'}^{k_1k_2}\text{Re}\Bigl(
        U_{j-k'-k_1+1,p}U_{j-k-k_2+1,p}U^{\ast}_{j-k'-k_2+1,p'}\times
        &\\H^{r\ast}_{p',p}U^{\ast}_{j-k-k_1+1,p''}H^{s\ast}_{p'',p} 
        -U_{j-k'-k_1+1,p}U^{\ast}_{j-k-k_1+1,p}U^{\ast}_{j-k'-k_2+1,p'}H^{r\ast}_{p',p}U_{j-k-k_2+1,p''}\times  
        &\\H^{s}_{p'',p}-U_{j-k'-k_1+1,p}U^{\ast}_{j-k'-k_2+1,p}U^{\ast}_{j-k-k_1+1,p'}H^{r\ast}_{p',p}U_{j-k-k_2+1,p''}H^{s}_{p'',p}  \bigr)\Bigr).
    \end{aligned}
\end{equation}
Therefore, the local extremality of the solution $U$, for $\mathcal{B}_t(U)$ function, could be verified by its matrix elements $U_{ij}$.

\section{Wigner D-Matrices}\label{Appendix E}

Let $J_x, J_y, J_z$ be components of the \textit{angular momentum operator} $\mathbf{J}$. The three operators satisfy the following commutation relations,
\begin{equation}
[ J_x ,J_y ]= i J_z \quad
[ J_z ,J_x ]= i J_y \quad
[ J_y ,J_z ]= i J_x ,
\end{equation}
where the reduced Planck's constant is set to identity, $\hbar =1$. Mathematically, the operators $J_x, J_y, J_z$  generate the Lie algebra $\mathfrak{su}(2)$. The sum of squares of $J_x, J_y, J_z$  is known as \textit{the Casimir operator}, 
\begin{equation}
\mathbf{J}^2 =J_x^2+ J_y^2+ J_z^2
\end{equation}
which commutes with $J_x, J_y$ and $J_z$ operators. In particular, $\mathbf{J}^2$ might be diagonalized together with $J_z$, which defines an orthonormal basis of joint eigenvectors labeled by quantum numbers $j,m$,
\begin{align*}
\mathbf{J}^2 \ket{j,m} &= j(j+1)\ket{j,m}, \\
J_z\ket{j,m} &= m\ket{j,m}, 
\end{align*}
with $j=0, \tfrac{1}{2},1,\ldots$ and $m=-j,-j+1,\ldots,j$. Note that for a given $j$, the operator $J_z$ is non-degenerated and has exactly $2j+1$ distinct eigenvalues. 

A three-dimensional \textit{rotation operator} has the form
\begin{equation}
\mathcal{R} (\alpha,\beta,\gamma ) = e^{-i (\alpha J_z+\beta J_y+\gamma J_z)}.
\end{equation}
The \textit{Wigner D-matrix} \cite{Wigner} is a unitary matrix of dimension $2j+1$ defined in the angular momentum basis as
\begin{equation}
D^j_{m m'} (\alpha,\beta,\gamma ) := \bra{j,m'} \mathcal{R} (\alpha,\beta,\gamma ) \ket{j,m}.
\end{equation}
Recall, that the matrix elements of the operator $J_z$ in the angular momentum basis read
\begin{equation}
\bra{j,m'} J_y \ket{j,m} = \frac{1}{2i}  \Big[ \delta_{m',m+1} \sqrt{(j-m)(j+m+1)}
-\delta_{m',m-1} \sqrt{(j-m)(j-m+1)} \Big],
\end{equation}
then, the precise form of the Wigner D-matrix for $j=\tfrac{1}{2}$ is
\begin{equation}
D^{\tfrac{1}{2}}  (\alpha,\beta,\gamma ) =
\left(\begin{matrix}
\text{cos} \frac{\beta}{2} e^{-\tfrac{i}{2} (\alpha+\gamma )}
&&
-\text{sin} \frac{\beta}{2} e^{-\tfrac{i}{2} (\alpha -\gamma )}
\\
\text{sin} \frac{\theta}{2} e^{\tfrac{i}{2} (\alpha -\gamma )}
&&
\text{cos} \frac{\theta}{2} e^{\tfrac{i}{2} (\alpha+\gamma )}
\end{matrix}\right)
\end{equation}
and for $j=1$:
\begin{equation}
D^{1}  (\alpha,\beta,\gamma ) =
\left(\begin{matrix} \cos^2{\frac{\beta}{2}}e^{-i(\alpha+\gamma)}&-\sqrt{2}\cos{\frac{\beta}{2}}\sin{\frac{\beta}{2}}e^{-i\alpha}&\sin^2{\frac{\beta}{2}}e^{-i(\alpha-\gamma)}\\\sqrt{2}\cos{\frac{\beta}{2}}\sin{\frac{\beta}{2}}e^{-i\gamma}&\cos{\beta}&-\sqrt{2}\cos{\frac{\beta}{2}}\sin{\frac{\beta}{2}}e^{i\gamma}\\\sin^2{\frac{\beta}{2}}e^{i(\alpha-\gamma)}&\sqrt{2}\cos{\frac{\beta}{2}}\sin{\frac{\beta}{2}}e^{i\alpha}&\cos^2{\frac{\beta}{2}}e^{-i(\alpha+\gamma)}\end{matrix}\right).
\end{equation}
Wigner matrices $D^j$ represent rotations in the space $\mathcal{H}_N$ with $N=2j+1$, which allow us to generate from a single fiducial vector $\ket{\psi_1}$ the entire iso-coherent basis $\{\ket{\psi_1},\ldots,\ket{\psi_N}\}$, --~see also \cref{Appendix F}.

\section{Invariance under rotation}\label{Appendix F}

The Majorana representation presents a spin-$j$ state as a collection of $2j$ stars on a sphere. In this section, we discuss the behavior of such a representation, while rotating the sphere. 
Firstly, we show that the rotation of a sphere preserves the scalar products of related states, in particular, transforming any orthonormal basis into another orthonormal basis. Secondly, we show that the rotation of a sphere preserves the anticoherence (\ref{anticoherence}) of related states. Lastly, we present the rotation of a sphere in terms of the transformation of related spin-$j$ and symmetric states.  As a consequence, all orthonormal bases of extreme quantumness specified in this work are defined up to the global rotation of related Majorana stars, or equivalently, up to the action of any Wigner D-matrix on related spin-$j$ basis elements. 

Recall that there is a direct correspondence between a unitary evolution of a two-level system and rotations of the related Bloch sphere representation, which follows directly from the fact that the group SU(2) doubly covers SO(3). In particular, any two-dimensional unitary matrix is of the form $U=\left(\begin{smallmatrix}a & b\\ -b^\ast & a^\ast \end{smallmatrix}\right)$ with $|a|^2+|b|^2=1$ and hence can be expressed as
\begin{equation}
\label{Unitaryrepr}
U (\alpha, \beta ,\gamma ) =\left(\begin{matrix}
\text{cos} \frac{\beta}{2} e^{-\tfrac{i}{2} (\alpha+\gamma )}
&&
-\text{sin} \frac{\beta}{2} e^{-\tfrac{i}{2} (\alpha -\gamma )}
\\
\text{sin} \frac{\theta}{2} e^{\tfrac{i}{2} (\alpha -\gamma )}
&&
\text{cos} \frac{\theta}{2} e^{\tfrac{i}{2} (\alpha+\gamma )}
\end{matrix}\right)
.
\end{equation}
As a quantum state $\ket{\psi}\in \mathcal{H}_2$  can be presented as a single point on the Bloch sphere, its evolution under the above unitary operator might be presented as a sequence of three Euler rotations acting on the Bloch sphere. More precisely, rotation $\mathcal{R}_Z ( \alpha )$ of $\alpha$ angle along $Z$ axis, then $\mathcal{R}_X ( \beta)$ by $\beta$ angle along $X$ axis and $\mathcal{R}_Z ( \gamma )$ by $\gamma$ angle along $Z$ axis again on the Bloch sphere. We shall denote such a rotation by $\mathcal{R}(\alpha, \beta ,\gamma )$. 

Any symmetric state $\ket{\psi}\in \mathcal{H}_2^{\otimes 2j}$ remains symmetric under any action of the joint unitary local operator $U^{\otimes 2j }$ \cite{Ribeiro}. In fact, the reverse statement is also true, all local unitary operations that preserve a given symmetric state are of the form $U^{\otimes 2j }$ \cite{Mathonet}. For a unitary matrix represented in the form (\ref{Unitaryrepr}) the action of $U^{\otimes 2j }$ is equivalent to the rotation $\mathcal{R}(\alpha, \beta ,\gamma )$ of the sphere. As a consequence, any rotation of stars in the Bloch representation does not change the scalar product of underlying symmetric states $\ket{\psi_{1,2}}\in\mathcal{H}_2^{\otimes 2j}$ as $\bra{\psi_1} (U^{\otimes 2j })^\dagger \;U^{\otimes 2j } \ket{\psi_2} =
\braket{\psi_1 |\psi_2}$. In particular, the orthonormal basis is mapped to another orthonormal basis of symmetric states. Furthermore, the purity of the reduced density operator (\ref{EqAuxi}) does not change under local unitary transformation $U^{\otimes 2j }$, and hence is the same for states related by the rotation of a sphere in the Majorana representation.

Recall that the system of symmetric states of $2j$ qubits is related to the spin-$j$ system by isomorphism $F$, --~see Eq.~\cref{Identification}, hence it preserves a scalar product, i.e $\braket{\psi_3 |\psi_4}= \braket{F(\psi_3) |F(\psi_4)}$, where $\ket{\psi_{3,4}}\in\mathcal{H}_{2j+1}$. Moreover, by the definition, both states $\ket{\psi}$ and $\ket{F(\psi )} )$ have the same Majorana representation. As a consequence, any rotation of stars in the Bloch representation of the spin-$j$ system does not change the scalar product of underlying states, as was the case for symmetric states. Notice also that any rotation of the sphere does not change the coherence properties of the represented states. Indeed, for any symmetric state $\ket{F(\psi)}$ the purity of the reduced density operator (\ref{EqAuxi}) does not change under local unitary transformation $U^{\otimes 2j }$, or rotation of the sphere in the Majorana representation, respectively. Hence, both states, the initial one and the rotated one, achieve the same $t$-anticoherence (\ref{anticoherence}). We can summarize this discussion by the following two observations.

\begin{observation}
An action of a Wigner D-matrix $D^j(\alpha, \beta ,\gamma )$ on a collection of spin-$j$ sets $\{\ket{\psi_k}\}^m_{k=1}$ belonging to $\mathcal{H}_{2j+1}$ preserves their mutual scalar products and $t$-anticoherence measure $\mathcal{A}_t (\ket{\psi_k} )$. 
Furthermore, it is represented by the rotation $\mathcal{R}(\alpha, \beta ,\gamma )$ in the stellar representation, i.e.
\begin{equation}
\mathcal{M}  \Big( D^j(\alpha, \beta ,\gamma ) \ket{\psi_k} \Big)
=
\mathcal{R}(\alpha, \beta ,\gamma )
\mathcal{M} (\ket{\psi_k} ).
\end{equation}
\end{observation}

\begin{observation}
An action of a joint local unitary operator $U(\alpha, \beta ,\gamma )^{\otimes 2j }$ on a collection of symmetric states $\{\ket{\psi_{sym}^k}\}^m_{k=1}$ belonging to $\mathcal{H}^{\otimes 2j}_2$ preserves their mutual scalar products and purity. 
It is represented by the rotation $\mathcal{R}(\alpha, \beta ,\gamma )$ in the stellar representation, i.e.
\begin{equation}
\mathcal{M}  \Big( U(\alpha, \beta ,\gamma )^{\otimes 2j } \ket{\psi_{sym}^k} \Big)
=
\mathcal{R}(\alpha, \beta ,\gamma )
\mathcal{M} (\ket{\psi_{sym}^k} ).
\end{equation}
\end{observation}

\newpage
\twocolumngrid

\end{document}